\documentclass[a4paper,fleqn,usenatbib]{mnras}

\usepackage{newtxtext,newtxmath}

\usepackage[T1]{fontenc}
\usepackage{ae,aecompl}

\usepackage{graphicx}	
\usepackage{url}        
\usepackage{multirow}   
\usepackage{amssymb}	
\usepackage{textcomp}   
\usepackage{caption}    
\usepackage{deluxetable}
\usepackage{hyperref}   
\usepackage{arydshln}

\newcommand{\Hersc}{{\it Herschel}}

\newcommand{\paperI}{Paper I}

\title[JINGLE: SCUBA-2 Data Release]{JINGLE, a JCMT legacy survey of dust and gas for galaxy evolution studies: II. SCUBA-2 850\micron\ data reduction and dust flux density catalogues}

\author[M. W. L. Smith et al.]{\Large \parbox{\textwidth}{Matthew W. L. Smith,$^{1}$\thanks{E-mail: Matthew.Smith@astro.cf.ac.uk}
Christopher~J.~R. Clark$^{1,2}$,
Ilse De Looze$^{3,4}$,
Isabella Lamperti$^{3}$,
Am\'elie Saintonge$^{3}$,
Christine~D. Wilson$^{5}$,
Gioacchino Accurso$^{3}$,
Elias Brinks$^{6}$,
Martin Bureau$^{7,8}$, 
Eun Jung Chung$^{9}$,
Phillip~J. Cigan$^{1}$, 
David~L. Clements$^{10}$, 
Thavisha Dharmawardena$^{11,12}$, 
Lapo Fanciullo$^{11}$,
Yang Gao$^{13}$,
Yu Gao$^{14}$,
Walter~K. Gear$^{1}$,
Haley~L. Gomez$^{1}$,
Joshua Greenslade$^{10}$,
Ho Seong Hwang$^{9,15}$,
Francisca Kemper$^{16,11}$,
Jong Chul Lee$^{9}$, 
Cheng Li$^{17}$,
Lihwai Lin$^{11}$, 
Lijie Liu$^{7}$, 
D\'{a}niel~Cs. Moln\'{a}r$^{18,19}$,
Angus Mok$^{20}$,
Hsi-An Pan$^{11}$,
Mark Sargent$^{18}$,
Peter Scicluna$^{11}$,
Connor~M.~A. Smith$^{1}$,
Sheona Urquhart$^{21}$,
Thomas~G. Williams$^{1}$,
Ting Xiao$^{13}$, 
Chentao Yang$^{14,22,23}$, 
\& Ming Zhu$^{24}$}
\\
\\
$^{1}$School of Physics and Astronomy, Cardiff University, Queens Buildings, The Parade, Cardiff, CF24 3AA, UK\\
$^{2}$Space Telescope Science Institute, 3700 San Martin Drive, Baltimore, Maryland, 21218, USA\\
$^{3}$Department of Physics \& Astronomy, University College London, Gower Street, London, WC1E 6BT, UK\\
$^{4}$Sterrenkundig Observatorium, Ghent University, Krijgslaan 281 - S9, 9000 Gent, Belgium\\
$^{5}$Department of Physics \& Astronomy, McMaster University, Hamilton, ON L8S 4M1 Canada\\
$^{6}$Centre for Astrophysics Research, University of Hertfordshire, College Lane AL10 9AB, UK\\
$^{7}$Sub-department of Astrophysics, University of Oxford, Denys Wilkinson Building, Keble Road, Oxford, OX1 3RH, UK\\
$^{8}$Yonsei Frontier Lab and Department of Astronomy, Yonsei University, 50 Yonsei-ro, Seodaemun-gu, Seoul 03722, Republic of Korea\\
$^{9}$Korea Astronomy and Space Science Institute, 776 Daedeokdae-ro, Yuseong-gu, Daejeon, 34055, South Korea\\
$^{10}$Blackett Laboratory, Physics Department, Imperial College, London, SW7 2AZ, UK \\
$^{11}$Academia Sinica, Institute of Astronomy and Astrophysics, No. 1, Sec. 4, Roosevelt Rd., 10617, Taipei, Taiwan \\
$^{12}$National Central University, No. 300, Zhongda Rd., Zhongli District, Taoyuan City 32001, Taiwan\\
$^{13}$Shanghai Astronomical Observatory, 80 Nandan Road, Xuhui District, Shanghai, China 200030 \\
$^{14}$Purple Mountain Observatory, Chinese Academy of Sciences, Nanjing 210034, China\\
$^{15}$School of Physics, Korea Institute for Advanced Study, 85 Hoegiro, Dongdaemun-gu, Seoul 02455, South Korea \\
$^{16}$European Southern Observatory, Karl-Schwarzchild-Str. 2, 85748 Garching b. M{\"{u}}nchen, Germany \\
$^{17}$Tsinghua Center for Astrophysics and Physics Department, Tsinghua University, Beijing 100084, China \\
$^{18}$Astronomy Centre, Department of Physics and Astronomy, University of Sussex, Brighton BN1 9QH, England \\
$^{19}$INAF - Osservatorio Astronomico di Cagliari, Via della Scienza 5, 09047 Selargius (CA), Italy\\
$^{20}$Department of Physics \& Astronomy, University of Toledo, Toledo, OH 43606, USA\\
$^{21}$School of Physical Sciences, The Open University, Walton Hall, Milton Keynes MK7 6AA, UK\\
$^{22}$Key Laboratory of Radio Astronomy, Chinese Academy of Sciences, Nanjing 210008, China\\
$^{23}$European Southern Observatory, Alonso de C{\'o}rdova 3107, Casilla 19001, Victacura, Santiago, Chile\\
$^{24}$National Astronomical Observatory of China, 20A Datun Road, Chaoyang District, Beijing, China 100012 \\
}

\date{Accepted 2019 April 13. Received 2019 April 12; in original form 2019 March 19.}

\pubyear{2018}
\begin{document}
\label{firstpage}
\pagerange{\pageref{firstpage}--\pageref{lastpage}}
\maketitle

\begin{abstract}
We present the SCUBA-2 850\micron\ component of JINGLE, the new JCMT large survey for dust and gas in nearby galaxies,
which with 193 galaxies is the largest targeted survey of nearby galaxies at 850\micron. 
We provide details of our SCUBA-2 data reduction pipeline, optimised for slightly extended sources, and including a calibration model adjusted to match conventions used in other far-infrared data. We measure total integrated fluxes for the entire JINGLE sample 
in 10 infrared/submillimetre bands, including all WISE, \Hersc-PACS, \Hersc-SPIRE and SCUBA-2 850\micron\ maps, 
statistically accounting for the contamination by 
CO($J$=3-2) in the 850\micron\ band. Of our initial sample of 193 galaxies, 191 are detected at 250\micron\ with a $\geq 5\sigma$ significance. In the SCUBA-2 850\micron\ band we detect 126 galaxies with $\geq 3\sigma$ significance. The distribution of the JINGLE galaxies in 
far-infrared/sub-millimetre colour-colour plots reveals that the sample is not well fit by single modified-blackbody models that assume a single dust-emissivity index ($\beta$). Instead, our new 850\micron\ data suggest either that a large fraction of our objects require $\beta<1.5$, or that a model allowing for an excess of sub-mm emission (e.g., a broken dust emissivity law, or a very cold dust component $\lesssim$10\,K) is required. We provide relations to convert far-infrared colours to dust temperature and $\beta$ for JINGLE-like galaxies. For JINGLE the FIR colours
correlate more strongly with star-formation rate surface-density rather than the stellar surface-density, suggesting
heating of dust is greater due to younger rather than older stellar-populations, consistent with the low proportion of
early-type galaxies in the sample.
 
\end{abstract}




\section{Introduction}

Studies of the interstellar medium in large, varied galaxy samples are crucial to our understanding of star formation and galaxy evolution. 
Surveys of both atomic gas (via the H{\sc i} 21cm line) and molecular gas (often traced by emission lines of  
the CO molecule) have revealed that there are key scaling relations in the local universe between global galaxy 
properties and the contents of the interstellar medium (ISM) \citep[e.g.][]{roberts94, young95, catinella10, Saintonge2011, Bothwell2013, Boselli2014b}. For example the Schmidt-Kennicutt
law \citep{schmidt59,kennicutt89} relates the surface density of star-formation to the surface density of gas in the galaxy. Large studies using H{\sc i}, dust continuum emission in the far-infrared (FIR), CO and other molecular line tracers have revealed that these scaling laws can depend on factors such as morphological type, mass, and environment \citep[e.g.][]{young11, lisenfeld11, Cortese2011,Smith2012a,Cortese2012, Clark2015,deVis2017,Remy-Ruyer2014}. 

Studies of dust in the ISM are important as over the lifetime of the Universe half of all light emitted from stars has been absorbed by dust and then re-emitted in the far-infrared \citep{Lagache2005}. Stars are formed in dense clouds of gas and dust, and so far-infrared observations can be vital for measuring an accurate star-formation rate in galaxies due to absorption of the UV and optical light \citep{Kennicutt1998, Calzetti2001}. Dust is important for molecules in the ISM as it catalyses reactions as atoms bind to the surface of dust grains \citep[e.g.][]{gould63,hagen79,vandishoeck04}. Given the difficulty with directly measuring molecular gas with CO or other tracers, and this especially at high redshifts, dust is also seen as a promising probe of the entire cold ISM \citep[e.g.][]{Guelin1993,Guelin1995,israel97, scoville14}. 

While there have been surveys with observations of H{\sc i}, CO and dust continuum, they are the exceptions. 
One example is the SINGS sample \citep{Kennicutt2003} which targeted $\sim$70 galaxies with distances $<30$\,Mpc,  
obtaining dust continuum data with \Hersc\ \citep[KINGFISH,][]{Kennicutt2011}, and exquisite gas measurements in H{\sc i} 
\citep[THINGS,][]{Walter2008} and CO \citep[HERACLES,][]{Leroy2009}. Another example is the \Hersc\ Reference Survey 
\citep[HRS,][]{Boselli2010b} which targeted 322 $K$-band selected galaxies in a volume-limited sample, and has collected
data on all three components of the cold ISM \citep{Ciesla2012, Cortese2014, Boselli2014a}.
To make sure to sample a full range of galaxy properties, large statistical samples beyond the
very local Universe focusing on these different components of the cold ISM are required. 
Using the James Clark Maxwell Telescope (JCMT), the JCMT dust and gas In Nearby Galaxies Legacy Exploration \citep[JINGLE;][hereafter \paperI]{JINGLE1}, aims to address this. JINGLE has observed dust continuum at 850\micron\ for 193 SDSS-selected galaxies ($M_{\ast}>10^9 {\rm M_{\odot}}$), and CO(\textit{J}=2-1) line emission for a subset of $\sim$35\% of them. The sample was selected in fields with coverage from {\it Herschel-}ATLAS \citep[hereafter \textit{H}-ATLAS,][]{Eales2010}, which observes between 100-500\micron,
the Arecibo ALFALFA HI survey \citep{giovanelli05}, and the MaNGA and SAMI optical integral field spectroscopic surveys \citep{bundy15,bryant15}, providing additionally the all important information about dust, atomic gas, 
ionised gas, and resolved stellar properties.

The SCUBA-2 data (the subject of this paper) of 193 galaxies is the largest targeted survey of nearby 
galaxies at 850\micron. Adding a data point far down the Rayleigh--Jeans tail
provides an improvement on obtaining dust measurements, over just using \Hersc\ data between 
70--500\micron. For example, there can be a degeneracy in fitted parameters 
between the dust emissivity index ($\beta$, see Section~\ref{sec:FIRdustBB}) and dust temperature measured \citep{Smith2012b}. By
adding a data-point at 850\micron\ further down the Rayleigh-Jeans tail of blackbody emission the degeneracy
can be reduced by approximately a factor of two. An additional benefit is that the \Hersc\ SPIRE calibration
is dominated by correlated uncertainties (4\% correlated versus 1.5\% uncorrelated) 
between bands due to uncertainties in planet models \citep{Bendo2013}, which can 
have a significant effect on SED fitting results. With an independent longer wavelength point at 850\micron\
(as SCUBA-2 uses different calibrators) we can obtain a more accurate measurement of $\beta$, which is important as
it can be an indication of grain size, composition, or surface mantles. Differences in the assumed numerical value of 
$\beta$ can have a significant effect on the dust-mass absorption coefficient ($\kappa_d$) and consequently on the
dust masses that are inferred from far-infrared observations \citep[e.g.][]{Bianchi2013, Clark2016}.

In this paper we present our observing strategy (Section~\ref{sec:obsStrategy}) and our customised data reduction 
pipeline, which is optimised for the specific requirements of the JINGLE survey (see Section~\ref{sec:dataRed}). 
In Section~\ref{sec:photometry} we describe our flux extraction methods for both point and extended sources, and 
describe our simulations to account for any flux attenuation from the pipeline high-pass filter. 
Section~\ref{sec:COcontamination} describes our approach to remove CO(\textit{J}=3-2) contamination of our 850\micron\
fluxes. Finally, in Section~\ref{sec:results} we investigate FIR/submm colour ratios, and how they relate to model fits of
temperature, dust emissivity, and other physical properties of the galaxy.

\section{Sample and observations}

\subsection{JINGLE sample overview}
The sources in the JINGLE survey are selected based on detections from the \textit{H}-ATLAS survey \citep{Eales2010} which
observed $\sim$600 sq. degrees with \Hersc\ \citep{Pilbratt2010}. \textit{H}-ATLAS observed in parallel mode 
using PACS at 100 and 160\micron, and SPIRE at 250, 350 and 500\micron\ simultaneously. As JCMT is in the northern 
hemisphere, JINGLE selected objects in the equatorial GAMA fields (161\,deg$^2$) and the North 
Galactic Pole (NGP) field (180.1\,deg$^2$). For our photometry (Section~\ref{sec:photometry}) 
we use the PACS maps provided in \textit{H}-ATLAS 
DR1 \citep[GAMA fields,][]{Valiante2016} and DR2 \citep[NGP,][]{Smith2017}. For SPIRE we use the same 
timelines used to generate the maps in the DR1 and DR2 releases, except we apply the relative gain 
corrections and calibration corrections to optimise the maps
for extended sources (these maps are also made available in the JINGLE data release).

\subsection{SCUBA-2 observations}
\label{sec:obsStrategy}
\begin{figure}
  \centering
  \includegraphics[trim=12mm 22mm 35mm 20mm, clip=True, width=0.49\textwidth]{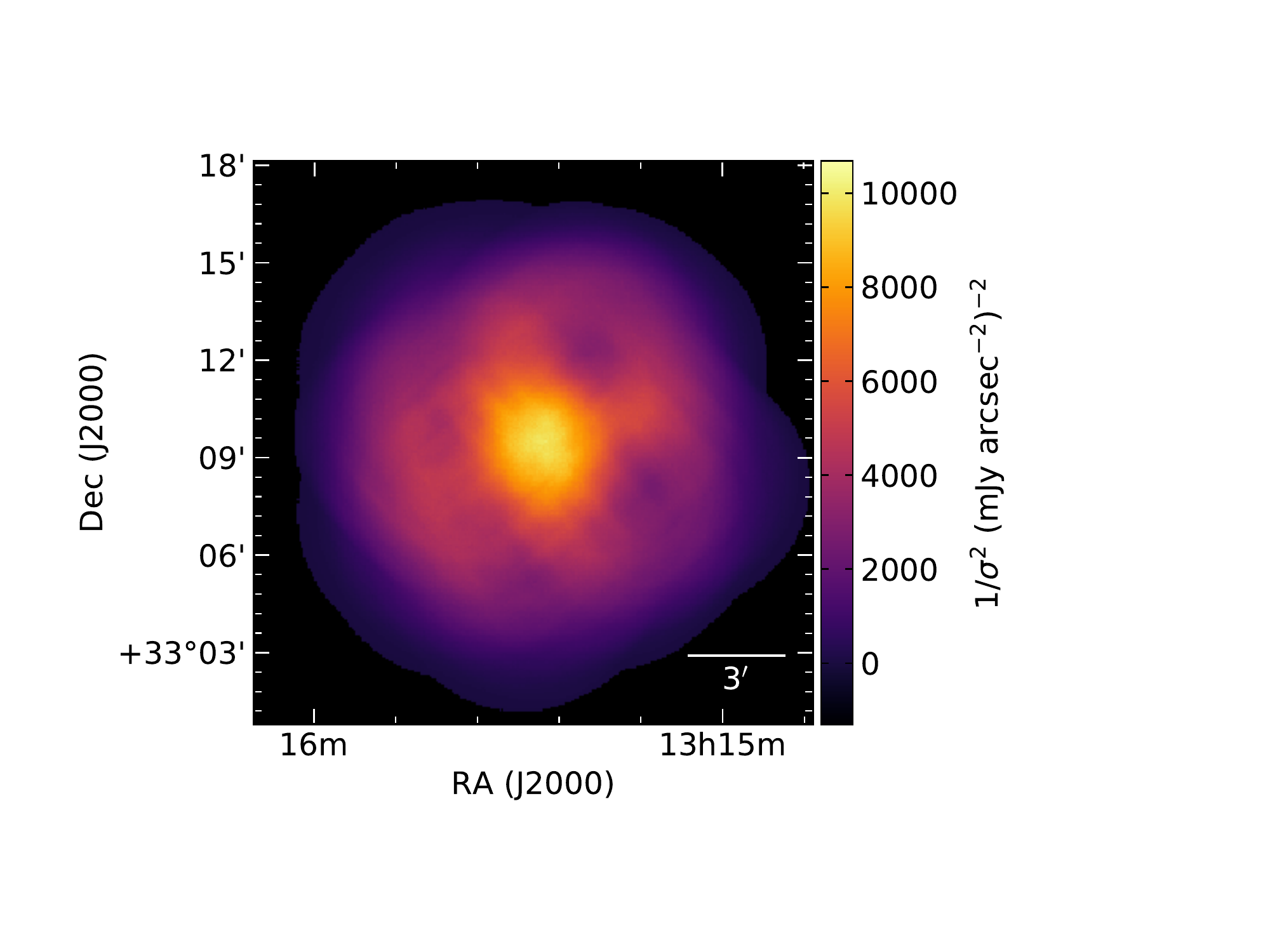}
  \caption{An example weight image for a typical observation, which illustrates how the sensitivity
           of our observation decreases with radius. The weight image is calculated from the inverse square of
           the noise.}
  \label{fig:weightImage}
\end{figure}

In this section we outline the strategy used by the JINGLE survey to obtain good quality maps
at 850\micron\ with SCUBA-2 (we also simultaneously obtain 450\micron\ but as the sensitivity is 
lower than SPIRE 500\micron\ we only concentrate on 850\micron\ in this work).
The redshift selection of the JINGLE survey ($0.01 < z< 0.05$) puts our targets at a distance 
of $\sim$41--212\,Mpc with a mean $D_{25}$ (the $B$-band isophotal diameter at a surface brightness of 25\,mag\,arcsec$^{-2}$)
of 1.28\arcmin, with the largest size of 4.6\arcmin. However, from the SPIRE 250\micron\ maps the largest aperture
required to accurately measure the flux of our objects has a diameter of 2.2\arcmin.
For all JINGLE targets we therefore observe using the `Constant Velocity (CV) Daisy' mapping mode which
is the smallest observing mode available with SCUBA-2, and provides an even coverage in the central 3\arcmin\ of the observation.
The `CV Daisy' is a circular scanning pattern designed so that the target is always within the
field-of-view of the array throughout the integration while moving at a constant 155\arcsec/s.
The observation provides useable coverage out to $\sim$6.0\arcmin\ in radius, but beyond
1.5\arcmin\ the map sensitivity decreases rapidly. A typical weight map ($1/\sigma^{2}$) 
for a JINGLE Daisy observation is shown in Figure~\ref{fig:weightImage}.

The sensitivity estimates for each JINGLE target were derived by fitting a 
modified-blackbody model to the \Hersc\ data (fluxes were taken from an initial $H$-ATLAS catalogue).
As literature values of the dust-emissivity index ($\beta$) tend to lie in the region 
1.5--2.0 \citep{Smith2012b, Planck2011, Cortese2014, Planck2014}, we 
use a constant value of $\beta = 2$, which should be a conservative estimate of the 850\micron\ flux.
To account for JINGLE galaxies being extended we use the radius of the galaxy to divide
the total flux evenly between 1--8 independent elements. The SCUBA-2 exposure time calculator was then used to calculate
the total integration time required to achieve a signal-to-noise ratio of 5.
The estimates were made assuming matched-filtering which as we discuss in Section~\ref{sec:dataRed} is not appropriate for our data, this is slightly compensated by our decision to use pixel sizes larger than the 1\arcsec\ default (when the survey was proposed) 
in the integration time calculator. The weather band (dependent on the amount of water vapour in the atmosphere) was chosen
so the total integration time is always under 2 hours. Overall our observations were generally taken in slightly better 
conditions than assumed when using the exposure time calculator.

\section{Data Reduction}
\label{sec:dataRed}

The JINGLE survey was designed specifically so as to minimise problems with emission on large-scales, which 
has been a problem with ground-based imaging due to atmospheric and instrumental variations often requiring
spatial filtering to be applied. To try avoid this issue the minimum redshift of $z > 0.01$ is used to select targets
without large angular sizes. However, by having isolated objects which are resolved, but do not 
have extended structures (like galactic regions or very-nearby galaxies) means JINGLE is between the two extremes of 
point sources or preserving large-scale emission for which SCUBA-2 pipelines have been developed. We therefore,
have worked to customise the data reduction pipeline for our specific case of marginally resolved objects. In this 
section we describe our customised pipeline.

\subsection{SCUBA-2 Data Processing}
\label{sec:map-making}

To reduce the raw SCUBA-2 data and create maps, we use the default SCUBA-2 map-maker {\sc MakeMap}
provided as part of the {\sc starlink} software package \citep{Currie2014}\footnote{All maps were created using {\sc Starlink} package version 2017A}.
We refer to \citet{Chapin2013} for full details of the {\sc MakeMap} algorithm, but briefly,
after an initial `cleaning' 
stage, which removes bad bolometers and artefacts such as glitches, 
{\sc MakeMap} begins an iterative procedure to split the bolometer signals into various components. 
In each iteration a `common-mode' signal predominately from sky noise is identified and subtracted, 
an extinction correction is applied, and then 
the bolometer signals are high-pass filtered. 
From these timelines (time-ordered detector readouts) a map is made to identify astronomical emission (AST model) which
is subtracted from the next iteration. This iterative procedure continues until the map converges (see below), or reaches the maximum
number of iterations. This technique is highly-customisable, allowing settings at every step to be adjusted to optimise the
data for your science target. Below we describe how we optimised the JINGLE data for our case of slightly extended
sources.

The standard {\sc MakeMap} implementation processes each observation individually and then combines the individual
maps at the end. If the memory requirements are too big for a machine, the observation is split into 
`chunks', each processed separately (note for all JINGLE processing a machine with 
enough memory is used so no `chunking' is required). 
While this is computationally efficient it does not make best use of the data, instead the astronomical
model in each iteration should be estimated from all the observations due to improved sensitivity and better resilience
to atmospheric/instrumental variations. The {\sc Skyloop} script provided in {\sc Starlink} solves this problem,
and is designed to help recover more signal, by using all the data 
at the end of every iteration. Our script is modified from the standard {\sc Skyloop} as we found that for all but the 
first observation passed to {\sc MakeMap}, the proportion of data flagged became excessively high. 
We modified the script to output a map for each observation after every iteration and then mosaic the data, instead 
of feeding all the data to {\sc MakeMap} in one go. After discussions with the observatory, {\sc Skyloop} in the
{\sc Starlink} 2017A package has been updated so this problem no longer exists.

The second difference is our method to apply source masking to the data. 
In SCUBA-2 terminology `Masking' is where regions are selected where the AST model is allowed to vary,
in any other area of the observation the AST model is set to zero. This improves the reduction as it helps
to reduce any degeneracies between AST and other noise models.
Such masks can be static (i.e., fixed for all iterations) or dynamic where the mask
is generated based on some threshold in the map at each configuration. For JINGLE due to the high-sensitivity of \Hersc\
SPIRE observations we have very good prior knowledge of where we expect to find emission at 450 and 850\micron, and 
can be confident we can select regions without missing any sub-millimetre emission. This masking
is very effective for our targets as our galaxies are isolated nearby sources and we therefore use the surrounding areas
as background. Unlike other surveys which apply a two-pass approach, where they perform 
an initial run to create a fixed mask by reducing the data either without a mask or with an adaptive mask,
through the use of our SPIRE data we can use our fixed mask. 
We generate a 2D-image as an input mask which has the advantage over the standard \textit{ast.zero\_circle} parameter in {\sc MakeMap}
that we can use elliptical shapes and include multiple sources. 
Our source ellipse is based on the size of the aperture used for our far-infrared photometry which we describe in detail in Section~\ref{sec:photometry}.

\begin{figure*}
  \centering
  \includegraphics[trim=2mm 13mm 8mm 19mm, clip=True, width=0.98\textwidth]{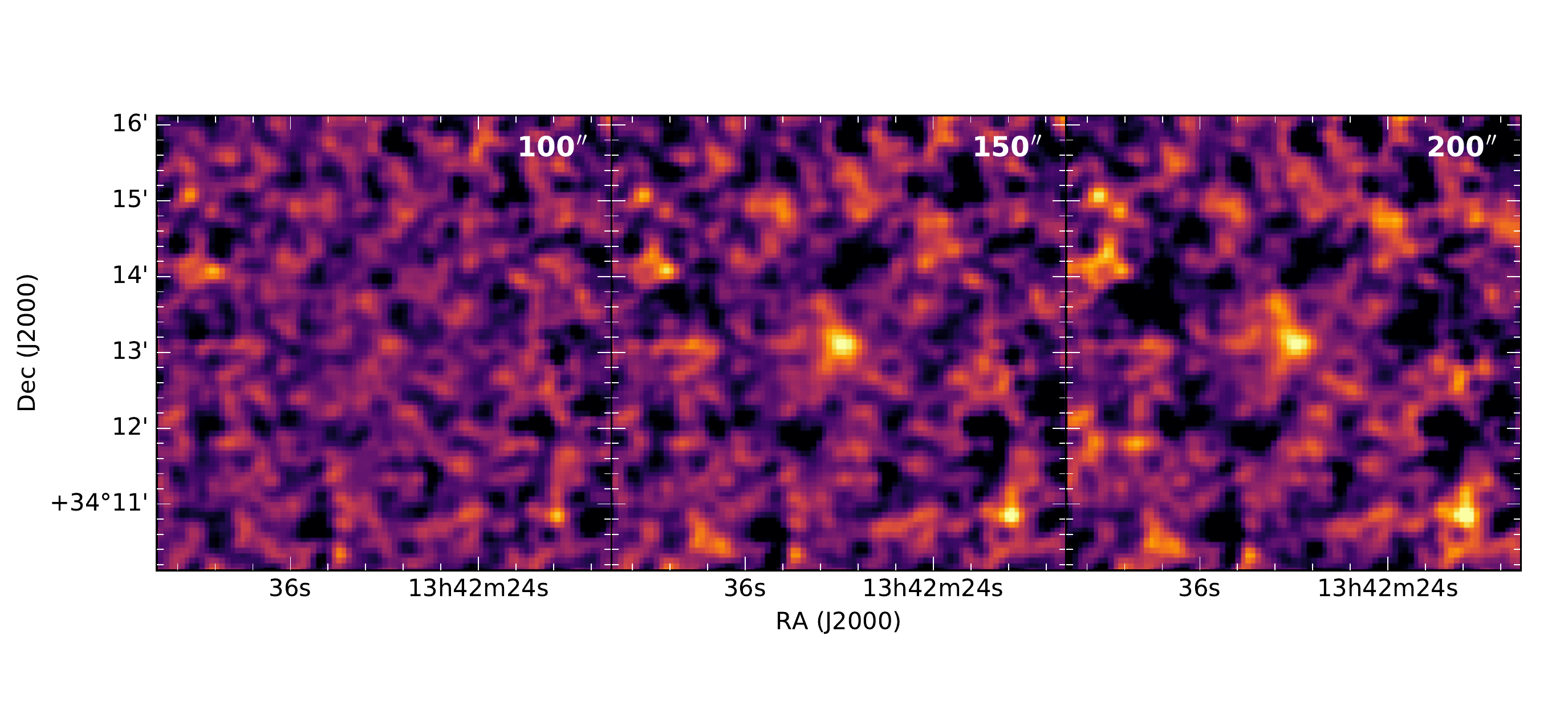}
  \caption{An example of how different filter scales (as set by the \textit{flt.filt\_edge\_largescale} parameter) 
           affect the resulting SCUBA-2 850\micron\ map for JINGLE~119. When a 100\arcsec\ filter (left) is applied our target is filtered out of the map, 
           but increasing the filtering to 200\arcsec\ (right) leads to increased noise.}
  \label{fig:filterScales}
\end{figure*}

The behaviour of {\sc MakeMap} is controlled using a configuration file which specifies all the settings to optimise
the map-maker for the science required. The JINGLE DR1 configuration file is made up of the following commands:
\begin{itemize}
 \item \textit{numiter = -300 \& maptol = 0.001} -- \citet{Mairs2015} found for 
 	   The JCMT Gould Belt Survey \citep{Ward-Thompson2007} that a
       stricter convergence tolerance helped recover more low-surface brightness structure in their maps. Whether a map has
       converged is set by the \textit{maptol} parameter which is the threshold value for the mean change of the map between iterations
       normalised by the root-mean-square of the pixel variances (i.e., when the map stops changing, note that this is calculated
       from regions not masked). The default of this parameter value is set to 0.05 and so the value of 0.001 is a much more
       stringent criterion. Like \citet{Mairs2015} to compensate for this stricter limit we increase the number of allowed iterations from 
       a typical value of $\sim$40 to 300 (the minus sign in the parameter allows the map to finish if converged, otherwise
       all iterations would be performed). We found that with our masking we normally converge reasonably 
       quickly ($\sim$20 iterations).

 \item \textit{com.perarray = 1} -- This parameter controls whether the `common-mode' component (i.e., atmosphere or
       instrument variations) is required to be the same across the whole array or whether each sub-array is 
       treated independently. We choose to set this so each sub-array is treated independently; this can remove emission scales
       above the size of a sub-array ($\sim$3\arcmin), but as we filter at smaller scales we can safely 
       treat each sub-array independently.
       
 \item \textit{flt.filt\_edge\_largescale = 150} -- This parameter has possibly the largest effect when creating SCUBA-2 maps 
 	   as it controls the scale (in arcseconds) of the high-pass filtering applied to the data. If set too high, 
 	   spurious signals from the atmosphere or drifts in the array could dominate the astronomical signal, however, 
 	   if set too low the filtering could remove a significant
       fraction of the source's emission. Figure~\ref{fig:filterScales} shows one 
       extreme example of the effect of the filter scale when we were testing our data reduction.
       When the filter scale was set to 100\arcsec, JINGLE~119 was not detected in our pipeline. 
       After experimenting with a test sample of objects we settled on a filter scale of 150\arcsec\ 
       as the best compromise between sensitivity and recovering the flux.
       In Section~\ref{sec:simulations} we describe our simulations to quantity the exact effect of the filtering on each
       source.
       
 \item \textit{ast.zero\_mask = 1 \& ast.zero\_snr = 0} -- These parameters set {\sc MakeMap} to use a static mask (as described above)
       using a file specified by the REF parameter, and to not apply 
       a threshold to the pixels within this mask to be included in the AST model.
       Specifying a file via the REF parameter sets the world coordinate system (WCS) for the output image to match the
       static mask. 
       
 \item \textit{ast.zero\_freeze = 0, com.zero\_freeze = 0 \& flt.zero\_freeze = 0} -- These are the default values which state
       that we do not freeze the AST, common-mode signal (COM), or low-frequency Fourier component (FLT), models after any iteration. These parameters are included as a previous version of
       skyloop required these parameters to be explicitly set.
 
\end{itemize}

We also investigated whether we should set the FLT mask using the same mask defined for the AST model 
(using the \textit{flt.zero\_mask} = 1. In the regions
set by the FLT mask the detector timelines are replaced by a linear interpolation before the high-pass filter is applied, 
with the aim of avoiding ringing around the source. By default this is only applied for the first two iterations and so the signal
can be identified in the AST model. We found that this had a large effect on our maps with our extracted flux densities being 
on average a factor of two higher than our predicted 850\micron\ fluxes, derived extrapolating modified blackbody fits to
\Hersc\ $\leq$500\micron\ flux densities. However, our standard reduction agreed well with our
predictions (this agrees with our simulations injecting sources into the map in Section~\ref{sec:simulations}). 
For some JINGLE targets setting the FLT mask resulted in an 850\micron\ flux density higher than the SPIRE 500\micron\
flux, which for a standard local galaxy (with no radio contamination) would be un-physical. As such
we do not set the FLT mask for our reduction, with the exception of JINGLE 70 and 132 which in our standard reduction resulted in
very significant negative fluxes (-4.7\,$\sigma$ and -7.5\,$\sigma$). Setting the FLT mask in these two cases removed the negative artefact and gave fluxes in line with our predictions.

\subsection{Calibration}

The standard procedure at the JCMT is to take a few calibration observations spread over the evening which can be used 
to see if there are any changes or particular problems over the night. JINGLE calibration observations are based
on either Mars, Uranus, CRL618, CRL2688 or Arp220 depending on source visibility. There are good reasons to assume the calibration 
does vary throughout the night, as the dish cools at the beginning of the night. As the pipeline is very flexible the standard 
advice is to apply the same procedure to your calibrators as adopted for the target object. For example, the default is to use a
30\arcsec\ radius aperture for your source and calibrator. However, while this advice is
sensible for point sources, it becomes problematic for extended sources where different fractions of the source will be
present in apertures, and filtering will have a different effect based on the amount being masked. We instead
decide to take an alternative approach, where we optimise the reduction of the calibrators to give the most accurate
calibration of our maps (i.e., the calibration that results in the true flux densities of our targets if no 
attenuation of large angular scales occurs), and then simulate the effect of filtering on our individual targets.

\begin{figure}
  \centering
  \includegraphics[trim=0mm 0mm 0mm 0mm, clip=True, width=0.49\textwidth]{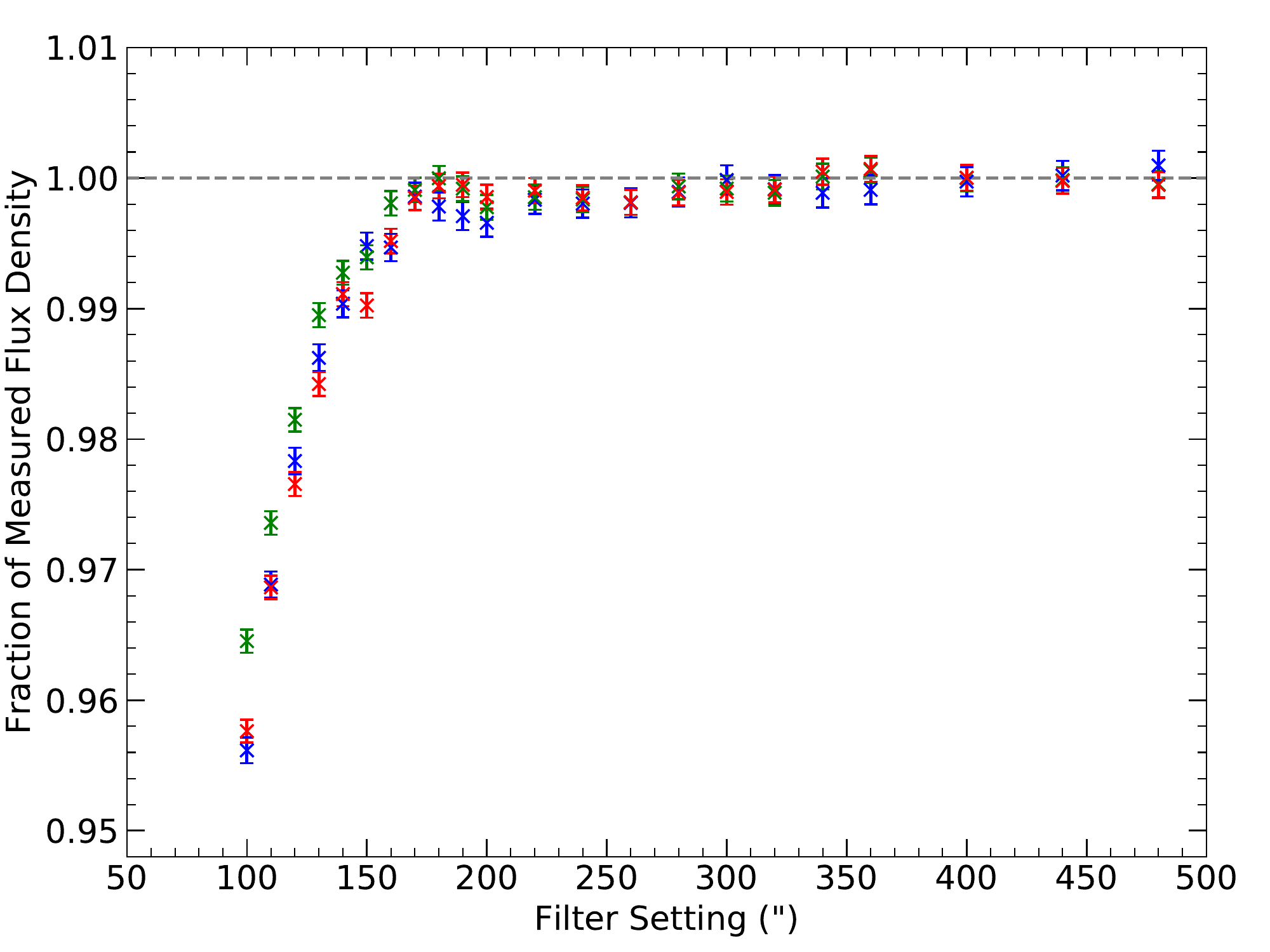}
  \caption{The fraction of flux measured compared to the average value measured in maps where the filter-scale has
           been set to $>$350\arcsec. The blue points correspond to observation id 20151226\_00069 with the
           com.perarray set to 0, the green and red points both correspond to observation id 20151227\_00067 
           with the com.perarray set to 0 (green) and 1 (red)}
  \label{fig:marsSim}
\end{figure}

To investigate the calibration we look at calibration observations from the beginning of the project to December 2016,
when we had observed just over half the SCUBA-2 sample.
When making maps of our calibrators, the aim was to use as similar as possible configuration file as used for our targets,
however, some adjustments were required. We discovered when creating maps of our calibrators that, in particular for 
Mars, the centre of the source was clipped due to the cleaning process. To mitigate this effect we set the
\textit{ast.mapspike} parameter to 10, which controls the signal-to-noise of detector samples to remove from a map-pixel, 
and the \textit{dcthresh} parameter which controls the signal-to-noise ratio of DC step detection to 10000.
As the calibration observations are centred on the calibrators, and the sources are all point-like in SCUBA-2 images  
we set the \textit{ast.zero\_circle} and \textit{flt.zero\_circle} to 2\arcmin, which sets the static mask 
(see Section~\ref{sec:map-making}) of the AST and FLT models to a circle in the centre of the image.
As with our target map-making in Section~\ref{sec:map-making}, one of the most important parameters 
is the filter-scale parameter (\textit{flt.filt\_edge\_largescale}). Ideally, the filter scale would be set to a 
very high value so the flux of our calibrator is not attenuated, however, as the calibration observation are short
for our dimmer calibrators this would lead to the measurement being dominated by large-scale noise on the map. To test
the optimum filter-scale so that our calibrators are not attenuated we take two independent observations of Mars 
(our brightest calibrator) and create a map for filter-scales from 100\arcsec\ to 200\arcsec\ in 10\arcsec\ intervals
and then from 200\arcsec\ to 400\arcsec\ in 20\arcsec\ intervals. Figure~\ref{fig:marsSim} shows the fraction of flux
measured depending on the filter-scale. From these results we chose to filter our calibrator observations on
a scale of 220\arcsec\ as this is safely on the plateau, but small enough to provide a flat background in the map, and
we set \textit{com.perarray} to 0.

\begin{figure*}
  \centering
  \includegraphics[trim=0mm 0mm 0mm 0mm, clip=True, width=0.80\textwidth]{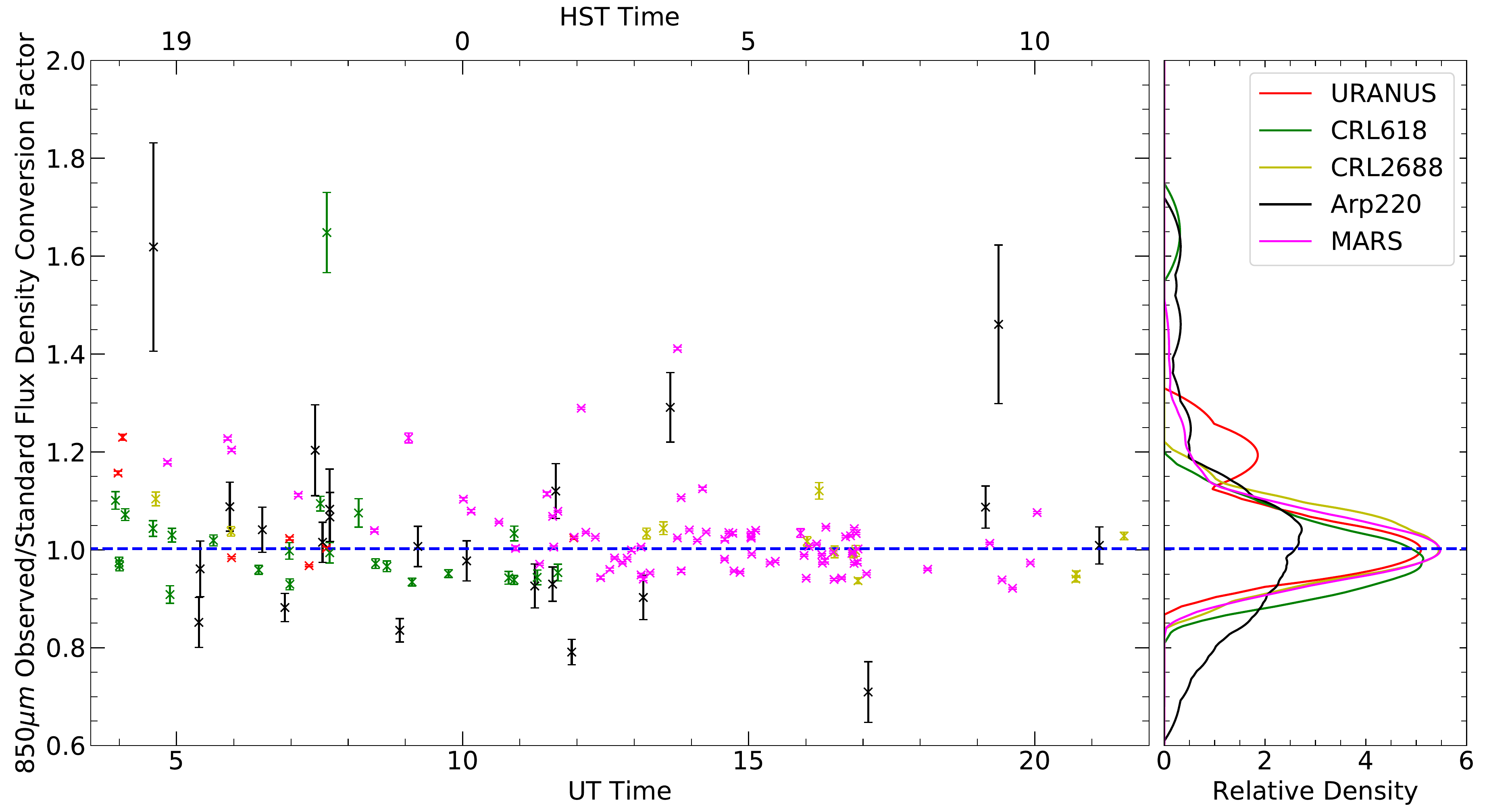}
  \caption{Flux conversion factors (FCFs) measured either before or after JINGLE observations at 850\micron\ up
           to December 2016, calculated
           from observations of Mars (magenta), Uranus (red), CRL618 (green), CRL2688 (yellow) and Arp220 (black).
           The blue line shows the median value of all our observed FCF values which is in very good agreement 
           with the standard FCF value (1.002$\times$). From the kernel density estimation (KDE) graph (right) 
           we see that most of the calibration sources agree well with each other, although Arp220 being a dimmer  
           source has a larger scatter. The data points are ordered by the time they were observed; we 
           see no obvious trends.}
  \label{fig:FCFvalue}
\end{figure*}

To calculate the flux conversion factor (FCF) from our calibrator maps we run the standard {\sc Picard} recipe
{\sc SCUBA2\_CHECK\_CAL} which uses the default settings which matches the method used by \citet{Dempsey2013} who
calculated the standard FCF values. As our sources are extended we calibrate our maps in units of mJy\,arcsec$^{-2}$
rather than mJy\,beam$^{-1}$. Figure~\ref{fig:FCFvalue} shows the ratio of flux conversion factors to the standard value 
obtained at 850\micron\ for our calibrators. The median ratio for our calibrators is 1.002 suggesting our observations
are consistent with the standard, although we find the scatter for Arp\,220 is significantly larger than the other sources,
probably due to it being one of the fainter calibrators. The scatter in FCF values result in a calibration uncertainty of 5.7\%,
similar to the value of 5\% found in \citet{Dempsey2013}.

The FCF values in Figure~\ref{fig:FCFvalue} are plotted against the time in the
day the observation was taken. We see no obvious trend to suggest the FCF is higher at the beginning of the night.
However, to check whether using a variable calibration leads to an improvement
over assuming a fixed calibration we reduce all maps using both methods. For the variable case we calculate the FCF 
for each observation based on the linear interpolation between the calibration observation before and after the
JINGLE observation (if observations are not available before and after we use the nearest calibration value). 
As {\sc Skyloop} takes all the data simultaneously we multiply the raw data for each science observation by the ratio of 
our variable FCF to standard FCF and then apply the standard FCF to the resultant map from {\sc Skyloop}. To see if there
is any improvement in the map we see if there is any reduction in the uncertainty map (accounting for the difference in
average FCF value). We did not see any maps with a significant improvement and so have decided to 
use our fixed values. We will re-visit this in subsequent works where the we use the 450\micron\ data, where the calibration
is more likely to vary.

The standard {\sc Picard} recipe to calculate the FCF uses a 30\arcsec\ radius aperture with a background annulus between
radii of 45--60\arcsec. While these apertures were chosen as a good compromise between encircling enough of the 
emission and achieving a good signal-to-noise, it systematically overestimates the FCF as it does not
account for the fraction of the beam outside the aperture. We decided to modify our FCF value to account for
this aperture correction, so our images match the calibration scheme of our other far-infrared datasets. 
To find this correction we use the measurements that
characterise the SCUBA-2 beam from \citet{Dempsey2013}, who find it is well described by the sum of two Gaussians,
the primary beam with a 13.0\arcsec\ FWHM and a secondary with a 48\arcsec\ FWHM. By integrating the beam profile
from a radius ($r$) of zero to infinity and within radii of 0--30\arcsec\ and 45--60\arcsec, we calculate 
what fraction of the beam is outside the standard aperture and the amount included within the background region. We
calculate that the FCF should be multiplied by a factor of 0.910 \citep[this is in reasonable agreement with 
values measured by,][]{Dempsey2013} and so we
apply an FCF of 2.134\,$\rm Jy\,pW^{-1}\,arcsec^{-2}$.

\subsection{WCS correction}

For a few objects we found cases where there appeared to be an offset between where we expected to see emission at
850\micron\ from images at other wavelengths (e.g., \Hersc). After investigating potential cases we found four cases
where small offsets seen in our observations were also seen in either the pointing or flux calibration observation 
performed after the science observations. From these observations we derive offsets of 3.91\arcsec, 
4.17\arcsec\ and 5.22\arcsec\ for JINGLE 35 (J131958.31+281449.3), JINGLE
149 (J125610.97+280947.4) and JINGLE 186 (J132035.40340821.7), respectively. For JINGLE 117,
observations on two separate days were found to have offsets of 6.84\arcsec\ and 4.43\arcsec, therefore 
these corrections are applied to the raw timeline data, before the map-making procedure.

\subsection{Individual Galaxy Considerations}

For the full JINGLE survey there are seven pairs of JINGLE targets and one triple system, 
where the JINGLE SCUBA-2 observations of each individual target overlap. As the full sensitivity region of 
Daisy maps is quite small and the only other map modes are significantly bigger and limited to circular observations,
we decided to observe these targets fully. These objects are processed together to make full use of the overlapping 
data, as the greater redundancy is useful for identifying and removing atmospheric and instrumental noise. 

For three galaxies (JINGLE 45, 119, 150) we found partially overlapping observations in the archive from M13AN02 \citep{Ivison2016} and M18AP013 (a JINGLE extension program), 
targeting other sources. For these galaxies combining the serendipitous observations with the JINGLE
data, leads to a $\sim$12, 5, and 26\% improvement in the instrumental noise for JINGLE 45, 119, 150, respectively.

We also inspected the SPIRE maps 
and our initial SCUBA-2 maps to see if there were other bright objects in the field
that should be included in the input mask. The serendipitously detected bright high-redshift objects described in
\paperI\ were also added to the mask and the maps re-run to ensure their flux was
not suppressed.

To test our choice of filter correction we also ran a reduction of the maps where we varied the filter scales to
a value of 175\arcsec\ for all galaxies and 200\arcsec\ for a selection of objects. 
For five galaxies we found a significant change in the flux measurement 
which were identified as cases in our 150\arcsec\ reduction where we find strong negative regions usually around a central 
source, often leading to a negative flux estimation. For JINGLE 45 we therefore use the map filtered at 
175\arcsec\ and for JINGLE 19, 44, 154 and 185, we use maps filtered at 200\arcsec.

\subsection*{}
The 193 SCUBA-2 maps described here, as well as the WISE, PACS and SPIRE cut-outs of our targets 
are available on the JINGLE data release page\footnote{\url{http://www.star.ucl.ac.uk/JINGLE/data.html}\label{url:JINGLEdr}}.
We also provide SCUBA-2 maps that have been Gaussian smoothed with either a FWHM of 12 and 24\arcsec, 
and a map generated by applying the {\sc Picard} matched-filtered algorithm 
which creates a map optimised for the detection of point sources. The code used to perform the data reduction is available
on GitHub\footnote{\url{https://github.com/mwls/SCUBA2-public}}.

\section{Far-Infrared/Sub-mm Flux Extraction}
\label{sec:photometry}

\subsection{Photometry}
\label{sec:apPhotomoetry}

Here we describe our method of deriving flux densities in the far-infrared and sub-mm (22--850\micron).
To extract flux densities from the \Hersc, \textit{WISE} and SCUBA-2 maps we perform aperture photometry
on the dust images separately from the procedures used at shorter wavelengths (i.e., UV, optical) 
presented in \paperI\ 
as our dust images (particularly in the SCUBA-2 bands) have reduced signal-to-noise, due to
the difficulties in observing in the far-infrared/sub-mm. Instead we decide to have a consistent
set of apertures optimised on the dust data, rather than the larger apertures that would be derived from optical 
wavelengths. These flux-densities are therefore ideal for the analysis of the 
dust properties of galaxies (i.e., far-infrared dust SED fitting).

The size and fluxes from our dust apertures are generated by an automated code, but a large degree of 
manual customisation is allowed to remove problems that arise (for example other bright objects or map artefacts).
To define our apertures we use the SPIRE 250\micron\ band, as at this wavelength we have the greatest 
sensitivity to dust structures. For the first step an automatic mask is created to remove the influence of 
any other bright nearby galaxy. We do this by performing a query on the RC3 catalogue \citep{Corwin1994} and masking all pixels
within a radius of 2.0\,$R_{25}$, which will safely mask the vast majority of dust contamination
\citep{Smith2016}. If other bright contaminating objects are 
found on the image their positions and sizes can be manually added to the mask. An initial noise estimate of the image is then
calculated using an iterative sigma-clipping technique. This rough noise is then used to find all 
contiguous pixels above a given signal-to-noise (set to a default of 3) within a region with $R < 0.5$\,$R_{25}$. 
The smallest ellipse which fits these pixels is then used to define the centre, 
the axes ratio, and the Position Angle (PA) of the ellipse. 
If no significant pixels exist, or the shape can not be found, we default to the
optical centre, the $D_{25}$ axis ratio, and PA, from the RC3 catalogue. Using these parameters, and our
rough noise estimates we then create a radial profile assuming a default background radius between 
1.1 and 1.4\,$R_{25}$. If the galaxy is above our detection criterion of a signal-to-noise of 3 then
the object is preliminarily classed as a detection and the preliminary size of the aperture is 
determined by the radius at which our radial surface-brightness ratio crosses a signal-to-noise of 2, and
multiplied by a default factor of 1.2 (this can be adjusted for special cases, i.e., to avoid
contaminating objects) to ensure we enclose most of the emission.

To calculate the final aperture we need to make a better estimate of the noise in the image.
To do this we use an estimate based on the method presented in \citet{Smith2013} and \citet{Ciesla2012}, 
where we consider the instrumental noise, confusion noise and background error separately. The instrumental noise is
found using the `error' extension provided with the maps, which are typically calculated based on
the standard deviation of samples contributing to that pixels. As a full map is available the exact
instrumental noise for a particular aperture is calculated by adding the pixels in the noise map in quadrature.
As the confusion noise is not independent between pixels (as pixels are smaller than the beam), the confusion
noise is calculated based on the square-root of the number of beams within the aperture. The third 
contribution is from large-scale backgrounds, including cirrus contamination. The method employed to measure
the large-scale background by \citet{Ciesla2012}
was very conservative and led to an overestimate of the uncertainty \citep[see ][]{Smith2013}. 
We therefore modify this approach and instead 
use {\sc Nebuliser}\footnote{\url{http://casu.ast.cam.ac.uk/surveys-projects/software-release/background-filtering}},
also used by the \textit{H}-ATLAS team \citep{Valiante2016},
which filters the map based on a threshold radius into small-scale and large-scale. We set this threshold 
to 90\arcsec\ in the SPIRE bands. We then
use our large-scale map (or effectively cirrus map) and place a series of aperture across the image with the same
size, shape and background region as our preliminary 
aperture on the cirrus map, and obtain the standard deviation of these apertures.
Using {\sc Nebuliser} in this way prevents us counting the same 
noise components multiple times. Ideally, these  
apertures would be randomly placed, but as some images
are limited in size (especially in the SCUBA-2 bands), and we are primarily interested in 
local conditions, we use a grid of apertures around the target source
(again avoiding other nearby galaxies). We scale this 
estimate for the relative number of pixels in our final aperture, which should be accurate assuming that the size 
of our preliminary aperture is similar to our final result.

As the point spread function (PSF) in far-infrared instruments can have significant extended features, we
apply aperture corrections. These can be quite large factors, as even though the
emission is low-surface brightness, the emission can be spread over a large area, and can
contaminate background regions leading to an over-subtraction of the flux. As dust is expected
to be distributed roughly in dust discs \citep{Pohlen2010,Hunt2015,Smith2016}, we fit 
an exponential plus constant model using the same PA and
axis-ratio as the pixels within the aperture. In the fitting process each trial model is convolved with 
the radially averaged beam. By using our model within the aperture, we then predict the amount of
emission outside the aperture, correcting our flux for the missing emission, and the effect of emission
in the background region. For bands with lower resolution we use the the model parameters found from the 
250\micron\ image and convolve the model with the appropriate beam
(see Table~\ref{tab:extractionParam} for whether the model was fit in each band).

\begin{figure*}
  \centering
  \includegraphics[trim=0mm 0mm 0mm 0mm, clip=True, width=0.99\textwidth]{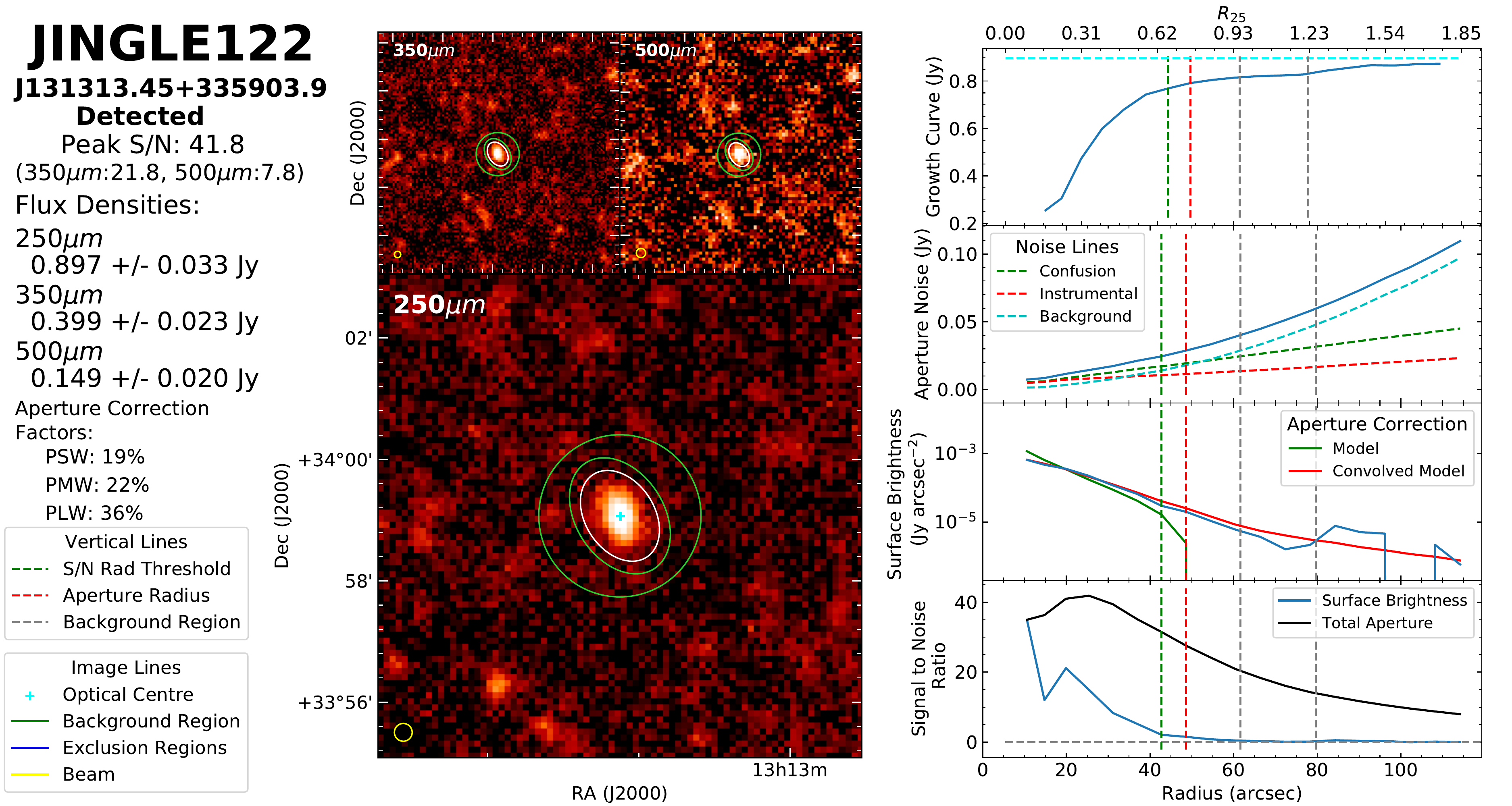}
  \caption{The diagnostic image produced by the photometry pipeline for JINGLE 122 for the SPIRE wavelengths.
           The images in the centre show the aperture (white) and the background region used (green) on the images, as
           well as the optical centre (cyan cross) and FWHM beam size in yellow. If objects were to be excluded from the 
           analysis they would be shown in blue. On the right of the graphic we have various curves which vary with radius;
           the top plot shows the growth curve (horizontal cyan line is the flux measurement), the second from top plot shows
           how the aperture noise grows as function of the different noise sources, third from top is the surface brightness
           profile (blue) with the exponential aperture correction models shown (green model, red convolved with beam), and 
           the bottom plot shows the signal-to-noise of the surface brightness profile (blue) and total aperture (black).
           On all the radial plots the vertical green line shows where the surface brightness profile crosses a signal-to-noise
           of 2 and the red line is the radius of the aperture. The vertical grey lines show the radii of the background region.
           Finally on the left the peak signal-to-noise, fluxes, and aperture corrections are listed.}
  \label{fig:SPIREphotometryPlot}
\end{figure*}

\begin{figure*}
  \centering
  \includegraphics[trim=0mm 0mm 0mm 0mm, clip=True, width=0.99\textwidth]{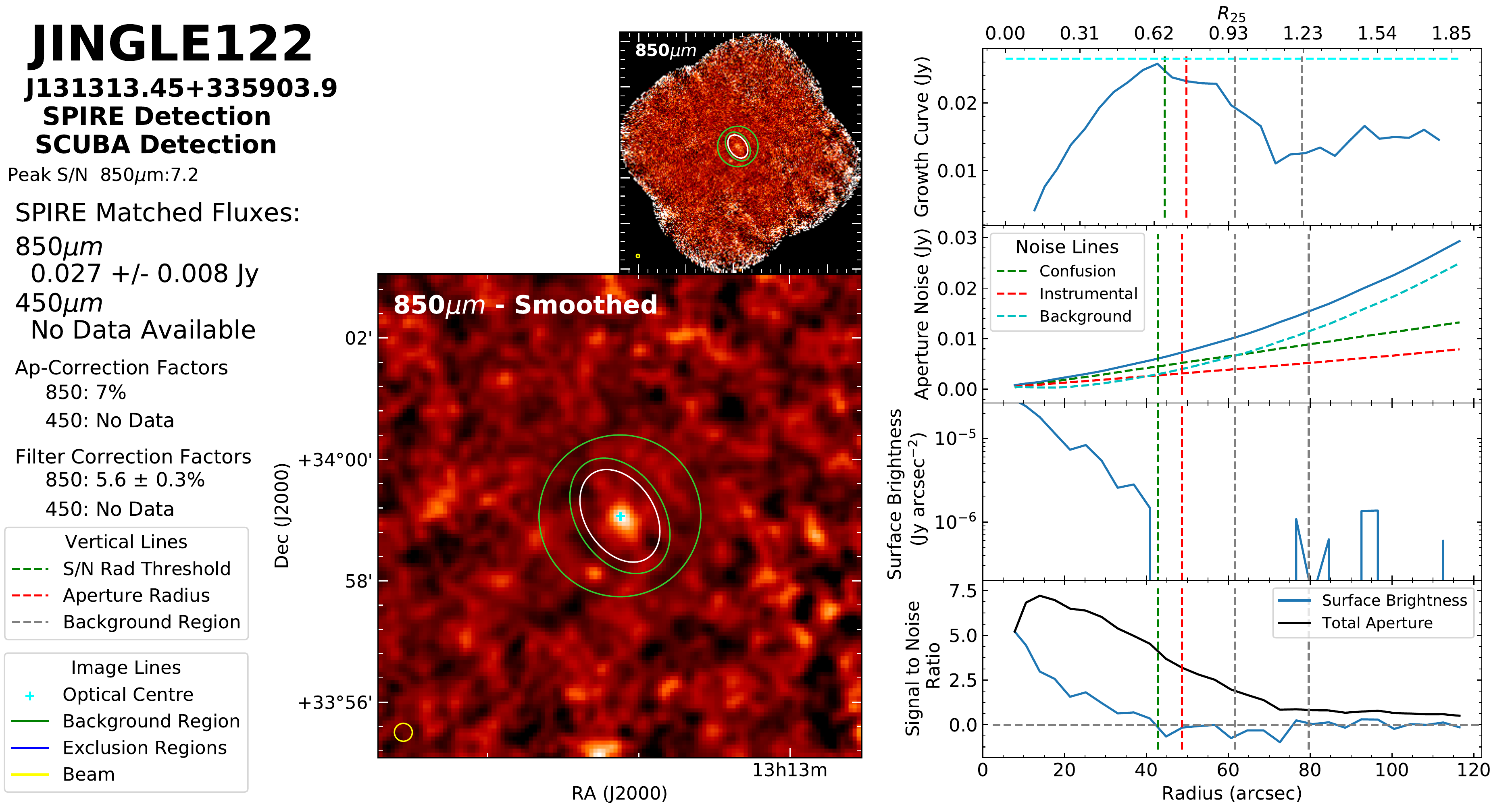}
  \caption{Same as Figure~\ref{fig:SPIREphotometryPlot}, except now applied to SCUBA-2. Unlike the SPIRE 250\micron\ figure
           the aperture is matched to the SPIRE band (rather than derived from the image).}
  \label{fig:SCUBA2photometryPlot}
\end{figure*}

\begin{table*}
\caption{Individual Band Parameters}
\begin{tabular}{cccccc}
\hline \hline
\multirow{2}{*}{Instrument} & \multirow{2}{*}{Band} & \multirow{2}{*}{FWHM} & Beam & Confusion & Nebuliser\\
                            &                       &                      & Area & Noise     & Median Filter\\
                            &                       & (\arcsec)            & (arcsecond$^2$) & (mJy\,beam$^{-1}$) & (\arcsec)\\
\hline
\multirow{3}{*}{SPIRE} & 250\micron\ & 17.6 & 469.7 & 5.8 & 90 \\
                       & 350\micron\ & 23.9 & 831.7 & 6.3 & 90 \\
                       & 500\micron\ & 35.2 & 1793.5 & 6.8 & 90 \\
\hdashline
\multirow{2}{*}{PACS}  & 100\micron\ & 11.4 & 147.2$^{\rm a}$ & 0.27 & 60 \\
                       & 160\micron\ & 13.7 & 212.7$^{\rm a}$ & 0.92 & 60 \\
\hdashline
\multirow{4}{*}{WISE}  & 3.4\micron\ & 6.1 & 42.2$^{\rm a}$ & - & 60 \\
                       & 4.6\micron\ & 6.4 & 46.4$^{\rm a}$ & - & 60 \\
                       &  12\micron\ & 6.5 & 47.9$^{\rm a}$ & - & 60 \\
                       &  22\micron\ &12.0 & 163.2$^{\rm a}$ & - & 60 \\
\hdashline
SCUBA-2                & 850\micron\ & 13.0 & 229.5 & ($m=1.22$ $c=-0.61$)$^{\rm b}$ & 60 \\
\hline
\end{tabular}\\
\begin{flushleft}
\textbf{Notes.} For the WISE wavelengths we assume the contribution of 
confusion noise is negligible and so set to zero in the code. Confusion noise estimates for SPIRE were taken from
\citet{Nguyen2010}, with the PACS FWHM, beam areas and confusion noise taken from \citet{Smith2017}.\\
$^{\rm a}$ These beam areas assume a Gaussian beam profile calculated from the FWHM. These instruments are not calibrated
in Jy\,beam$^{-1}$ (or similar), but the conversion is used just to scale the confusion noise (where appropriate).\\
$^{\rm b}$ See Section~\ref{sec:apPhotomoetry} for details on the SCUBA-2 confusion noise.
\end{flushleft}
\label{tab:extractionParam}
\end{table*}

While for the vast majority of objects in our sample the automated apertures are valid, a 
small subset (15) of objects require us to lock the centre to the optical centre. For 16 galaxies 
we lowered the default aperture expansion-factor of 1.2, 
although the factor is never set to be less than 1.0. These changes were often
made to avoid problems from nearby background sources or other contaminating features. 

In bands other than the SPIRE 250\micron, we use a matched aperture to the 250\micron\ band,
except we correct for the difference in beam size by modifying the size of the semi-major and semi-minor axis.
This is achieved by taking the 250\micron\ aperture size and subtracting in quadrature the 250\micron\ FWHM/2 and
adding the current band's FWHM/2. Finally a useful diagnostic graphic is created showing the images with apertures, the
resultant fluxes, peak signal-to-noise ratios, aperture corrections, the radial profile and growth curves. An example
of this image is shown for JINGLE 122 in both SPIRE (Figure~\ref{fig:SPIREphotometryPlot}) and 
SCUBA-2 (Figure~\ref{fig:SCUBA2photometryPlot}) wavebands.

Confusion measurements for SCUBA-2 are difficult as surveys of galaxies at higher redshifts 
apply matched-filtering to optimise the detection of point sources. This convolution changes the confusion noise 
measured on the map. To find the confusion noise we perform a simulation where we create a fake map that just 
contains confusion noise, and then apply the SCUBA-2 match filter. We vary the level of confusion until the amount
of confusion noise in the raw map matches the 0.8\,mJy\,beam$^{-1}$ measured by the SCUBA-2 Cosmology Legacy Survey 
\citep{Geach2017}. As part of our test we also check how the confusion noise scales with aperture size, we found that for SCUBA-2 
assuming the contribution of confusion varying as $\sigma_{conf} \sqrt{N_{\rm beam}}$ did not represent the noise adequately,
instead we use $\sigma_{conf} \sqrt{N_{\rm beam}} - const$ where $\sigma_{conf}$ is 1.22\,mJy\,beam$^{-1}$ and $const$ is -0.61.
Table~\ref{tab:extractionParam} lists the beam area, confusion noise, and nebuliser filter scale assumed for the
flux density measurement in each band.

Both JINGLE 57 and 60 at 850\micron\ were found to have significant contamination within the aperture from a point source 
presumably at a higher redshift. To remove the contamination we fit the point source using the same procedure outlined in
Section~\ref{sec:pointSource}, and then subtract the source from the image. We then run the aperture code as normal on the
subtracted image. A more complicated case is JINGLE 29 where the more compact early-type object overlaps with a more flocculent
late-type galaxy. To separate these objects we use the typical size of dust-disks based on the $D_{25}$ from \citet{Smith2012a}
to choose a region that likely encompasses the extent of emission from JINGLE 29, this also agreed with the region that would
be chosen visually. Then all pixels within that region contaminated by the late-type galaxy were 
replaced with a sigma-clipped mean of the map,
and the resulting image then used to perform the usual aperture flux extraction method.

\subsection{Point Source Extraction}
\label{sec:pointSource}

\begin{figure*}
  \centering
  \includegraphics[trim=12.5cm 0mm 12mm 0mm, clip=True, width=0.8\textwidth]{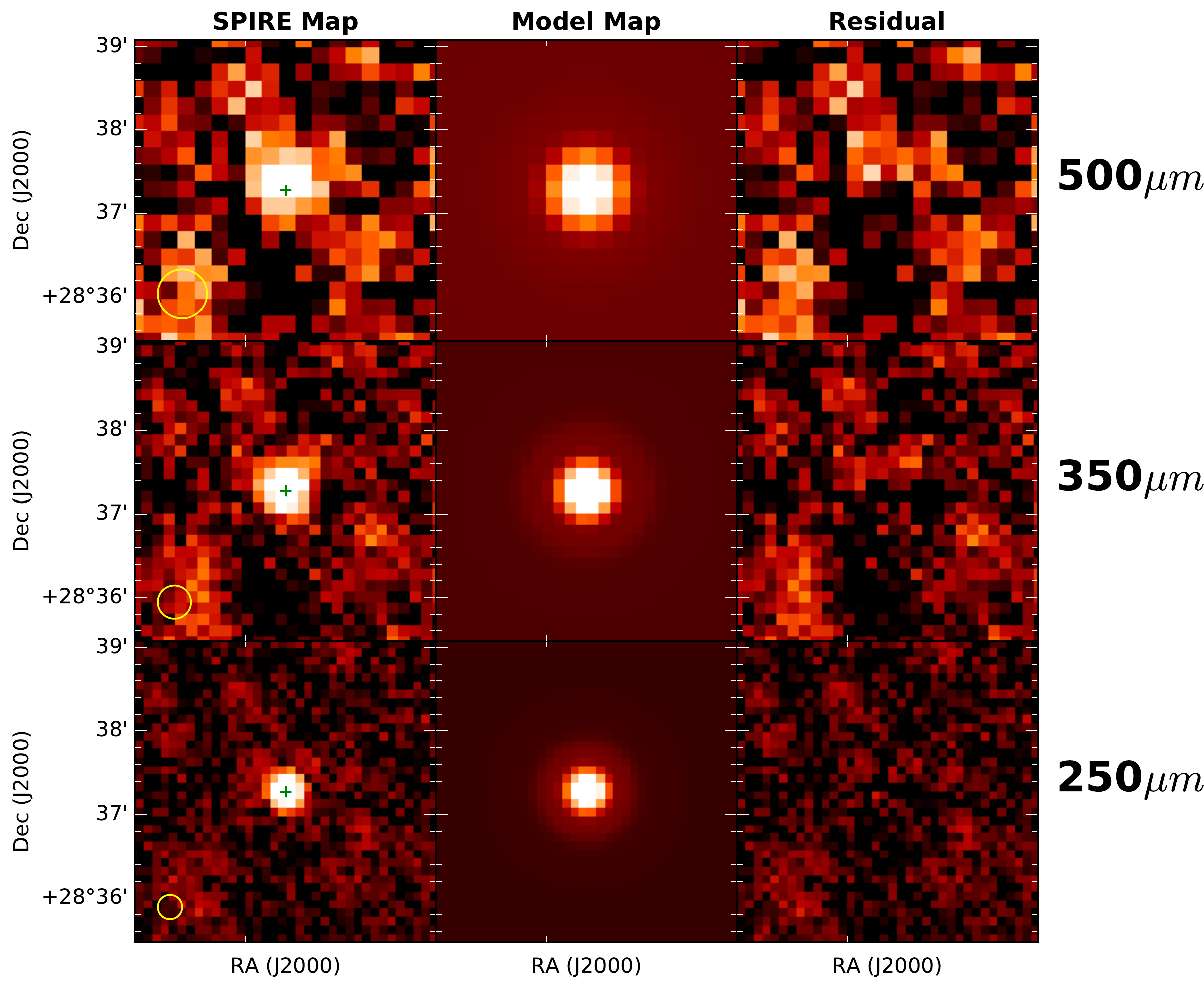}
  \caption{An example of the point source fitting method for JINGLE~33 in the SPIRE wavelengths.
           The grid of images show the SPIRE maps (left column), the point source model images (middle column),
           and the residual images after the point source is subtracted (right column) for the 250, 350 and 
           500\micron\ bands. On the SPIRE maps the optical centre is shown by the cyan cross, the fitted centre by the 
           green cross (in this example the two overlap so only the green cross is visible) and the beam FWHM in yellow.}
  \label{fig:SPIREpointPlot}
\end{figure*}

Several of our sources are compact enough they are well described by a point source in the far-infrared 
wavelengths from PACS to SCUBA-2. By identifying point sources we can use extraction techniques optimised
to recover flux densities using the PSF as a prior, rather than relying on aperture photometry. This improves
both the flux estimate and the significance of the detection (particularly in the SCUBA-2 bands).
For SPIRE to fit a point source the images are firstly converted into units of Jy\,beam$^{-1}$ and then {\sc Nebuliser} is
run over the maps to remove any emission on scales greater than 3\arcmin.
We then fit a radially averaged PSF in each band, using a model grid based on five times smaller pixels 
than the original pixel size to minimise the effect of pixelisation. For the SPIRE 250\micron\ we allow the central position to 
vary by up to 6\arcsec\ compared to the optical centre, while for other bands the central position is locked 
to the SPIRE 250\micron\ position. The largest offset was found to be 3.3\arcsec\ with most offsets substantially
smaller. For the SPIRE bands the uncertainty in the flux is measured using the uncertainties measured in the fit combined with
the confusion noise (as given in Table~\ref{tab:extractionParam}). 
An example of the point source extraction applied to a SPIRE image is shown in Figure~\ref{fig:SPIREpointPlot}.

To decide whether our candidate galaxies can be reasonably approximated by point sources as well as a visual 
inspection, we run the residual images
through our aperture photometry code as described in Section~\ref{sec:apPhotomoetry}. If the residual signal in
the aperture at 250\micron\ has a signal-to-noise $< 2.5$ then we use the point source procedure. From our sample of
193 galaxies, 42 meet our criteria to be considered point-like.

For the SCUBA-2 images we first apply the `match-filter' {\sc orac-dr} recipe which optimises SCUBA-2 maps for point source
extraction. Effectively the recipe subtracts a Gaussian smoothed version of the map
and convolves the map with the SCUBA-2 PSF. We then fit the matched-filter PSF characterised by \citet{Geach2017}, with the source centre
fixed to the SPIRE position. However, as the matched-filtering convolves the image the noise of each pixel can no 
longer be considered independent. We therefore perform a Monte Carlo simulation where for every pixel in the 
raw map we scale a normally distributed random number to account for the instrumental noise in the 
pixel and add it to the map. The same matched-filtering and point source extraction is performed to the new map 
and the whole process repeated 100 times. The uncertainty in the flux is then estimated based on the distribution
of extracted fluxes. Note, that despite the slightly higher angular resolution of the JCMT at 850\micron\ than SPIRE at 
250\micron, with the smoothing action of the matched-filter these sources should be well represented by a point source).

Three candidate point-galaxies that narrowly failed our criteria (JINGLE 162, 182 and 192) at SCUBA-2 850\micron, were
found to have higher flux measurements when using the point source measurement, compared to the aperture photometry
method, due to noise in the aperture. For these three galaxies we therefore report the point source flux (this again
is helped by the smoothing action of the matched-filter).

The \textit{H}-ATLAS team recommends for PACS data that apertures of radius 11.4 and 13.7\arcsec\ are used for the extraction
of point sources \citep{Valiante2016,Smith2017}, due to variations of the beam shape. As the WISE data is of similar or better
resolution we use the same 11.4\arcsec\ radius for the W1--W3 bands and 13.7\arcsec\ radius aperture for the W4 band.
Whether the point source extraction method is used in either the SPIRE or the SCUBA-2 850\micron\ bands is specified in the
flux-density catalogue table available in the supplementary materials or on the JINGLE data release page$^{\ref{url:JINGLEdr}}$.

\subsection{Simulations}
\label{sec:simulations}

As mentioned in Section~\ref{sec:map-making}, the SCUBA-2 pipeline relies on high-pass filtering, and while we have
tried to minimise the effect when creating a map there could be some residual flux attenuation. To try to correct
for this effect for each object we run a simulation where we inject a model of the source into the raw instrumental data
of JINGLE observations that appear to have no signal (although to avoid any bias we avoid the location of the target galaxy). 
From our observations observed up to December 2016 we identified 12 target regions that could be used for this purpose.
The raw data with the injected source is then processed with the same map-making procedure as for the real data 
(including same mask size and configuration). To minimise the
effect of the noise on our correction, we also re-run the map-making but with no model source injected using the same mask. This blank map
is then subtracted from the map with the injected source and estimates of both the aperture and filter corrections are calculated.

To make the model that is injected into the raw data we take the exponential models used to find the aperture corrections in the SPIRE
data (see Section~\ref{sec:apPhotomoetry}), these are then convolved with the SCUBA-2 beam and the flux adjusted to match our predicted
850\micron\ flux. The model source is then converted into pW so it matches the units of the instrumental timelines. For each JINGLE
target we inject the galaxy into three different blank fields chosen to have the closest sensitivity to our target observation 
(estimated in the central region of the instrumental uncertainty map). The model source to be injected is adjusted so the
signal-to-noise ratio is the same for the injected source as expected in the real map (from the predicted flux-density). The centre and 
orientation of the injected source is chosen randomly, but the distance from the centre of the image is limited to be 
within 2\arcmin\ so we are not affected by the noisy regions at the edges of the map. The random location is not allowed to overlap
with the intended target of the `blank' data. The size of the mask used is identical to that applied to the reduction of the 
target observation.

For galaxies that we assume are point-like the correction factor (this correction also accounts for any systematic from
the matched-filtering and flux extraction process we have not accounted for) has an average value of 1.135 with a narrow
distribution (minimum factor 1.121 and maximum 1.158). For extended galaxies the average filter correction 
is lower at 1.045, but has a much larger range from 1.006 to 1.139. These corrections have been applied to all fluxes 
provided in the catalogue.

\subsection{Catalogue Statistics}
\label{sec:catStats}

At 250\micron\ 191 of the 193 galaxies in the JINGLE sample are easily detected ($>5\sigma$) by 
SPIRE with most objects having peak signal-to-noise ratios between 10--50 (the high fraction 
is not surprising as galaxies were included in our sample based on a preliminary \textit{H}-ATLAS catalogue). 
The two notable exceptions are JINGLE 62 and JINGLE 130, neither of which are listed in the release version of the
\textit{H}-ATLAS DR2 catalogue \citep{Maddox2018}. From our measurements JINGLE 130 is identified with a peak signal-to-noise 
ratio of 4.5 and is identified as a point-like source. We therefore follow the same procedures as galaxies detected with
signal-to-noise ratios $>5$. JINGLE 62 is fainter than JINGLE 130 and has a significance far below 3.0\,$\sigma$, we therefore
just report an upper-limit to the flux for this source across all the wavebands, using an elliptical aperture with semi-major
axis equal to $R_{25}$ \citep[this size was chosen to match findings of the HRS,][]{Ciesla2012,Smith2012a}. 

Table~\ref{tab:detectionStats} provides an indication of our detection rates across the different bands, by giving the
number of objects with signal-to-noise ratios greater than 5 or 3. The table is split by peak signal-to-noise ratio, which
is the aperture that gives the highest signal-to-noise, while the total signal-to-noise is the measurement on the total
aperture. The peak measurement is more analogous to a detection statistic while the total aperture shows the ability to 
measure the total flux of an object (for point sources the peak and total S/N are the same). 
As outlined in Section~\ref{sec:photometry} we use the SPIRE 250\micron\ data to define our apertures, so that we get 
aperture matched fluxes (or point source estimates) across all wavebands even if the signal-to-noise ratio is significantly
lower in other bands.

\begin{table}
\caption{Sample Detection Statistics}
\begin{tabular}{cccccc}
\hline \hline
\multirow{2}{*}{Instrument} & \multirow{2}{*}{Band} & \multicolumn{2}{c}{Peak S/N} &  \multicolumn{2}{c}{Total S/N} \\
                            &                             & $\geq 5\sigma$ & $\geq 3\sigma$ & $\geq 5\sigma$ & $\geq 3\sigma$\\
\hline
\multirow{3}{*}{SPIRE} & 250\micron\ & 191/193 & 192/193 & 191/193 & 192/193 \\
                       & 350\micron\ & 186/193 & 190/193 & 185/193 & 190/193 \\
                       & 500\micron\ & \ 93/193 & 138/193 & \ 79/193 & 132/193 \\
\hdashline
\multirow{2}{*}{PACS}  & 100\micron\ & 177/190 & 181/190 & 157/190 & 174/190 \\
                       & 160\micron\ & 182/190 & 187/190 & 177/190 & 185/190 \\
\hdashline
\multirow{4}{*}{WISE}  & 3.4\micron\ & 192/193 & 192/193 & 189/193 & 189/193 \\
                       & 4.6\micron\ & 192/193 & 192/193 & 189/193 & 189/193 \\
                       &  12\micron\ & 192/193 & 192/193 & 192/193 & 192/193 \\
                       &  22\micron\ & 188/193 & 192/193 & 186/193 & 188/193 \\
\hdashline
SCUBA-2                & 850\micron\ & \ 62/193 & 126/193 & \ 30/193 & \ 83/193 \\
\hline
\end{tabular}\\
\begin{flushleft}
\textbf{Notes.} The total number of objects in the PACS sample is 190 instead of the 193 for SPIRE, WISE and SCUBA-2, as 3
galaxies were not covered by PACS imaging due to the
22\arcmin\ offset between the PACS and SPIRE cameras on the \Hersc\ focal plane.
\end{flushleft}
\label{tab:detectionStats}
\end{table}

\begin{figure}
  \centering
  \includegraphics[trim=11mm 4mm 0mm 0mm, clip=True, width=0.49\textwidth]{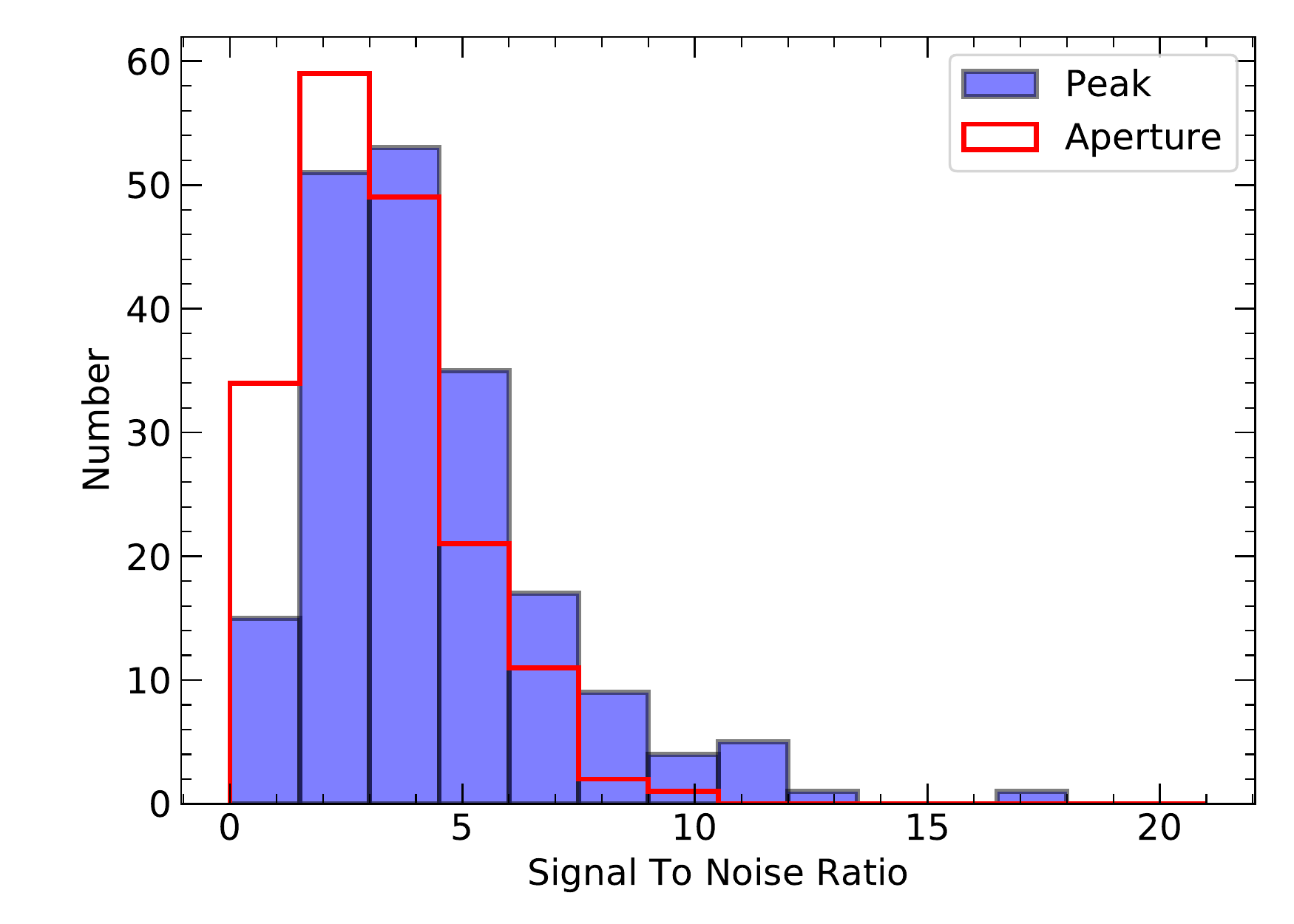}
  \caption{A histogram of the signal-to-noise ratio values for the flux measurements of JINGLE galaxies 
           at the SCUBA-2 850\micron\ wavelength. The blue filled histogram is the peak signal-to-noise 
           (see text) while the open red histogram is the signal-to-noise of the SPIRE matched aperture.}
  \label{fig:S2N-spireMat}
\end{figure}

Table~\ref{tab:detectionStats} shows that we have secure WISE and SPIRE 250/350\micron\ detections 
for almost every object in our sample and over 82\% with PACS coverage (with total S/N $>$ 5). At
SPIRE 500\micron\ as we move down the Rayleigh-Jeans tail the number of galaxies detected drops to 48\%
at 5\,$\sigma$, and 41\% have total fluxes with signal-to-noise ratios greater than 5. The JINGLE SCUBA-2
850\micron\ of course is more challenging due to the intrinsically fainter emission of local galaxies in this 
waveband and the effects of the 
atmosphere, but we do detect 126 of our 193 (65\%) at a level of $> 3\sigma$, and have 83 (43\%) galaxies where
the total signal-to-noise of the flux measurement is greater than 3. However, galaxies with lower signal-to-noise
can still be useful to constrain the dust properties of an object. The distribution of the peak and total aperture
signal-to-noise measurements for our SCUBA-2 sample is shown in Figure~\ref{fig:S2N-spireMat}.

\subsection*{}
The complete table of flux-densities in the WISE, PACS, SPIRE and the SCUBA-2 850\micron\ bands is available in the
supplementary materials and on the JINGLE data release page$^{\ref{url:JINGLEdr}}$. The python scripts used to perform the
photometry are available online\footnote{\url{https://github.com/mwls/Public-Scripts}}.

\section{CO(3-2) contamination}
\label{sec:COcontamination}

The CO($J$=3-2) emission line at 867.6\micron\ (354.796\,GHz) falls within the bandpass filter for the 
850\micron\ SCUBA-2 band for all the galaxies in the JINGLE sample ($z <\ 0.05$); leading to potentially significant 
contamination of the continuum flux \citep{Drabek2012}.
In order to correct for this contamination, we need to estimate the strength of the CO($J$=3-2) emission line. 
For the JINGLE galaxies we do not have CO($J$=3-2) observations, but we have predictions of the CO($J$=1-0) luminosity from 
Xiao et al. in prep (Paper III). In this paper, they show that these predictions are in good agreement with the CO line luminosities measured for a sub-sample of JINGLE galaxies.
The CO($J$=3-2) line luminosity can be estimated from the CO($J$=1-0) transition, 
assuming an excitation line ratio r$_{31} = L'_{\text{CO(J=3-2)}}/L'_{\text{CO(J=1-0)}}$.

To estimate the ratio r$_{31}$ for the JINGLE galaxies, we studied a sample of $\sim$20 galaxies from the COLD GASS survey
\citep{Saintonge2011} which allowed us to measure r$_{31}$ in a redshift range similar to that of the JINGLE survey.
The galaxies of this sample also have stellar masses, specific star-formation rates (SSFR) and star-formation efficiencies 
(SFE = SFR\,/\,M(H$_2$)) comparable to the JINGLE sample. We found a mean value of r$_{31} = 0.53 \pm 0.05$. 
This value is consistent with observations of low-redshift galaxies 
\citep[average r$_{31}$ = 0.6---0.7]{Yao2003, Mao2010, Papadopoulos2012} 
and intermediate redshift (z $\sim$ 0.3) star forming galaxies \citep[average r$_{31} = 0.46 \pm 0.07$]{Bauermeister2013}.
Therefore we decided to use the constant value r$_{31}$ = 0.5.

We converted the obtained line luminosity $L'_{\text{CO(J=3-2)}}$ into line intensity $I_{\text{CO(J=3-2)}}= \int T_{\text{MB}} \text{d}v$ in [K km s$^{-1}$] using eq.(2) from \citet{Solomon1997}:
\begin{equation}
I_{\text{CO(J=3-2)}} = \frac{L'_{\text{CO(J=3-2)}}}{23.5 \cdot \Omega_b D_L^2 (1+z)^{-3}}
\end{equation}
where $L'_{\text{CO(J=3-2)}}$ is in units of $\text{[K km s$^{-1}$ pc$^2$]}$, $\Omega_b$ is the telescope beam area in arcsec$^2$ and $D_L$ is the luminosity distance in Mpc. We assume that the sizes of our galaxies are comparable to the size of the SCUBA-2 beam (13.5" FWHM), and use $\Omega_b=\pi (13.5/2)^2$ arcsec$^2$.

To convert the line intensity from line units $\text{[K km s}^{-1}]$ to SCUBA-2 850\micron\ continuum flux, we used the conversion factor $C$ defined by \citet{Drabek2012} as:
\begin{equation}
C = \frac{F_{\nu}}{\int T_{\text{MB}} \text{d}v}=\frac{2k \nu^3}{c^3}\frac{g(\nu)}{\int g(\nu)\text{d}\nu}\Omega_B
\end{equation}
where $C$ has the units $\text{mJy beam}^{-1} \text{per K km s}^{-1}$.
$F_\nu$ is the line flux, $\int T_{\text{MB}} \text{d}v$ is the integrated main-beam temperature, $k$ is the Boltzmann constant, $g(\nu)$ is the transmission at the frequency $\nu$ of the 850\micron\ filter multiplied by the atmospheric transmission, and $\Omega_B$ is the telescope beam area.

The $C$ factor that converts the line intensity into SCUBA-2 continuum flux depends on the atmospheric transmission. 
We used the updated prescription for the $C$ conversion factor from \cite{Parsons2018}, which provides 
the $C$ factor as a function of the precipitable water vapour (PWV). The main differences from the 
prescription by \cite{Drabek2012} are that they used an updated main-beam FWHM of 13.0\arcsec\ with a 
relative amplitude of 0.98, and that they include a correction for the secondary beam component, which has 
a relative amplitude of 0.02 \citep{Dempsey2013}. The $C$ factor as a function of PWV is defined as:
\begin{multline}
C = 0.574 + 0.1151\cdot \text{PWV} - 0.0485\cdot \text{PWV}^2 + 0.0109\cdot \text{PWV}^3 \\
 - 0.000856\cdot \text{PWV}^4 \text{ mJy beam}^{-1}/\text{K km s}^{-1},
\end{multline}
where the PWV is related to the sky opacity at 225 GHz ($\tau_\text{225 GHz}$) as 
PWV = $(\tau_\text{225 GHz}-0.017)/0.04 \text{ mm}$. The $C$ factor for our sample varies 
in the range $0.63-0.75$ [mJy beam$^{-1}$ per K km s$^{-1}$]. 
The $C$ factors are calculated at the rest-frame frequency of CO($J$=3-2). To account for the different 
transmission at the observed frequency of CO($J$=3-2), we applied the following correction:
\begin{equation}
C_{obs} = \frac{1}{(1+z)^3}\cdot \frac{g(\nu_{obs})}{g(\nu_{rest})} \cdot C
\end{equation}
where $\nu_{rest}$ is the  CO($J$=3-2) rest-frame frequency and $\nu_{obs} = \nu_{rest}/(1+z)$ 
is the CO($J$=3-2) observed frequency. For each JINGLE observation we average the $\tau_{225 GHz}$ 
values for each observation and then calculate the correction.

Thus, the CO($J$=3-2) flux contamination in [mJy\,beam$^{-1}$] is:
\begin{equation}
F_{\text{CO(3-2)}} = \int T_{\text{MB}} \text{d}v \cdot C_{obs}
\end{equation}
The CO($J$=3-2) flux contamination is in the range 0.7--41.2\% of our predicted 850\micron\ values with a mean contamination of 10.1\%. 
For the vast majority of sources (78\%) the correction is less than 15\% of the predicted 850\micron\ flux density, and
91\% of sources have corrections less than 20\% of the predicted 850\micron\ flux density. The SCUBA-2 flux estimated contamination
is provided in the main catalogue table available in the Supplementary Materials. After correcting for the CO ($J$=3-2) flux contamination 73 of our targets have aperture fluxes with a signal-to-noise greater than 3.

\section{Results \& Discussion}
\label{sec:results}

\subsection{Comparison of the Predicted and Observed SCUBA-2 fluxes}

In this section, we compare the observed SCUBA-2 850\micron\ fluxes with predictions
from fitting the far-infrared SED using our \Hersc\ data 
in the wavelength range 100-500\micron. We used a single modified black-body model (SMBB) 
defined as \citep{Hildebrand1983}:
\begin{equation}
F_{\lambda} = \frac{M_{\text{dust}}}{D^{2}} 		\kappa_{\lambda} B_{\lambda}(T_{\text{dust}})
\end{equation}
where $M_{\text{dust}}$ is the dust mass in the galaxy, $D$ is the distance of the galaxy and $B_{\lambda}$($T_{\text{dust}}$) is the Planck function for the emission of a black-body with a dust temperature $T_{\text{dust}}$.
The dust mass absorption coefficient $\kappa$ varies as a function of wavelength:
\begin{equation}
 \kappa_{\lambda}= \kappa_{0}\left( \frac{\lambda_{0}}{\lambda}\right)^{\beta}
\end{equation} 
where $\kappa_{0}$ is the reference dust mass absorption coefficient. We use a constant value 
$\kappa_0 = 0.051$\,m$^2$\,kg$^{-1}$ at 500\micron\ from \citet{Clark2016}. We assume a fixed emissivity 
index $\beta=2$ to have conservative estimates of the flux, since $\beta$ is typically in the 
range 1.5--2.0 \citep[e.g,][]{Boselli2012, Galametz2012, Smith2012b, Clemens2013, Cortese2014}. We fit 
the model to the \Hersc\ fluxes (100, 160, 250, 350 and 500\micron), using a Bayesian non-hierarchical 
method implemented using the MCMC code {\tt emcee} \citep{Goodman2010,Foreman-Mackey2013}, and assuming 
that the noise is normally distributed. For details about the fitting procedure see Lamperti et al, in prep. 
The predicted spectra in the 850\micron\ band were convolved with the filter transmission curve of the SCUBA-2 band, 
before comparing them with the observed fluxes. 

\begin{figure*}
\centering
\includegraphics[trim=0mm 0mm 0mm 0mm, clip=True, width=0.9\textwidth]{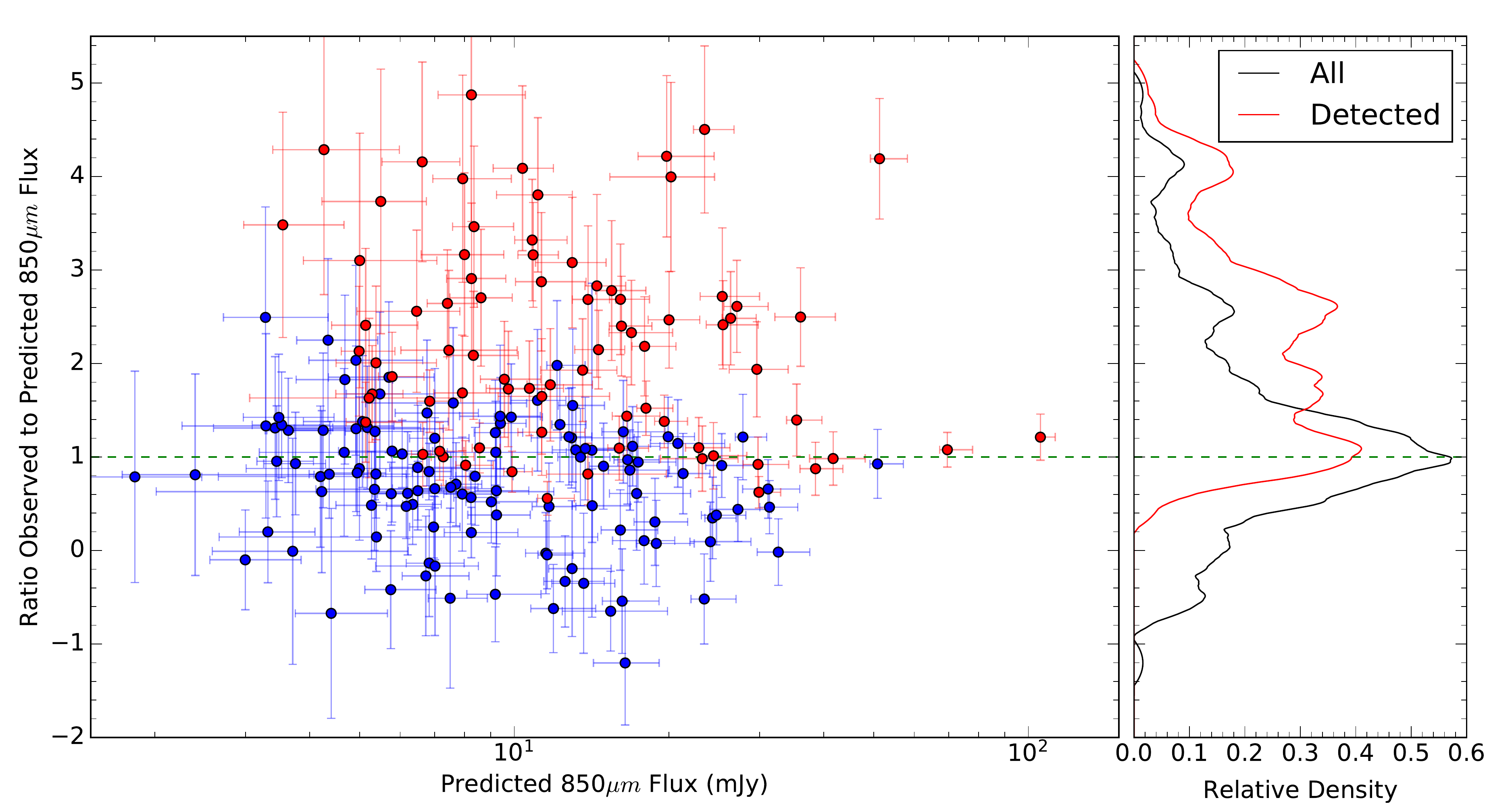}
\caption{The left panel shows the ratio of the observed flux to the predicted 850\micron\ flux versus 
         the predicted flux. The 850\micron\ flux predictions are based on a fit of a single modified 
         black-body to the \Hersc\ bands with a fixed emissivity index $\beta = 2$. Objects that have measured
         fluxes with signal-to-noise ratios greater than 3 are shown by the red points, and all other 
         objects in blue. The horizontal green dashed line is where the ratio of observed to predicted
         850\micron\ flux is 1. The right panel shows the KDE of the ratio of observed to predicted 850\micron\ flux
         for all objects (black line) and those with measured flux ratios with S/N $>$ 3 (red line).}
\label{fig:comp_F850_pred_obs} 
\end{figure*}

Figure~\ref{fig:comp_F850_pred_obs} shows the comparison between the predicted and observed 850\micron\ fluxes.
As the JINGLE observations were individually tuned to reach a fixed detection threshold (rather than a fixed sensitivity), galaxies with
a higher measured flux-density than predicted tend to be detections, while those with lower than predicted flux density do not reach our detection criteria.
While the kernel-density estimator (KDE) graph (right panel Figure~\ref{fig:comp_F850_pred_obs}), peaks in agreement
with our predictions, generally there is a large scatter. For the majority of objects ($\sim$75\%) the scatter 
could be explained simply due to the uncertainty in the flux measurement. However, the distribution is wider
than you would expect from the uncertainties (both above and below our predictions).
More galaxies (110 versus 82) show an 
excess of the observed fluxes (i.e., they are higher than the predictions), meaning that for those galaxies 
the slope of the SED in the sub-millimeter is shallower. The most likely explanation is that their dust SED should
be modelled with a lower $\beta$ value, either due to the intrinsic dust properties or
from multiple dust temperature components along the line of sight. 
Several studies have found $\beta$ varying between 0.5--2.5 on global scales 
\citep[e.g.,][]{Boselli2012, Galametz2012, Clemens2013, Cortese2014} and within a galaxy
\citep{Smith2012b,Tabatabaei2014,Kirkpatrick2014}. We will investigate this further in subsequent sections.

There are also galaxies for which the observed flux is lower than the predicted flux. Our predictions could overestimate
the true 850\micron\ flux density if the dust emissivity index 
could be greater than 2 \citep[e.g.,][]{Lis1998, Meny2007, Smith2012b}, or if we have overestimated the CO($J$=3-2) contribution
for these galaxies. For JINGLE we estimated the CO($J$=3-2) contribution from the predicted CO($J$=1-0) 
luminosity, using a constant line excitation correction factor $r_{31}= L_{CO(3-2)}/L_{CO(1-0)} =0.5$, 
but this factor is known to vary in the range 0.1--1.9 \citep{Mauersberger1999,Yao2003,Mao2010}. 
Assuming an extreme value of $r_{31}=0.1$, the CO($J$=3-2) contribution will be a factor of five lower. 
The estimated CO contribution is lower than 30\% of the observed flux for 95\% of our sample. 
Thus an overestimation of $r_{31}$  can only account for a deficit of $\sim$25\% of the observed 
flux with respect to the predictions. As shown in Figure \ref{fig:comp_F850_pred_obs}, the observed 
deficit can be as large as the predicted flux density, therefore the CO($J$=3-2) 
contribution is not sufficient to explain the deficit.

\subsection{Far-infrared Colours}

In this section we investigate how far-infrared/sub-mm colours vary with each other and compare
with the predictions of modified blackbody models. We define a FIR/sub-mm colour as the 
ratio of two flux-densities at wavelengths in the range 100--850\micron.
We investigate a selection of colour ratios that sample different regions of the far-infrared spectrum.
For example the ratio at smaller wavelengths like $F_{100}/F_{160}$, and $F_{100}/F_{250}$ should be more 
sensitive to temperature variations of the dust. Colour-ratios including flux densities at longer wavelengths
$F_{160}/F_{500}$, $F_{250}/F_{500}$, and $F_{160}/F_{850}$ should be more sensitive to changes in 
the dust emissivity index ($\beta$). 

\begin{figure*}
\centering
\includegraphics[trim=14mm 18mm 5mm 5mm, clip=True, width=0.95\textwidth]{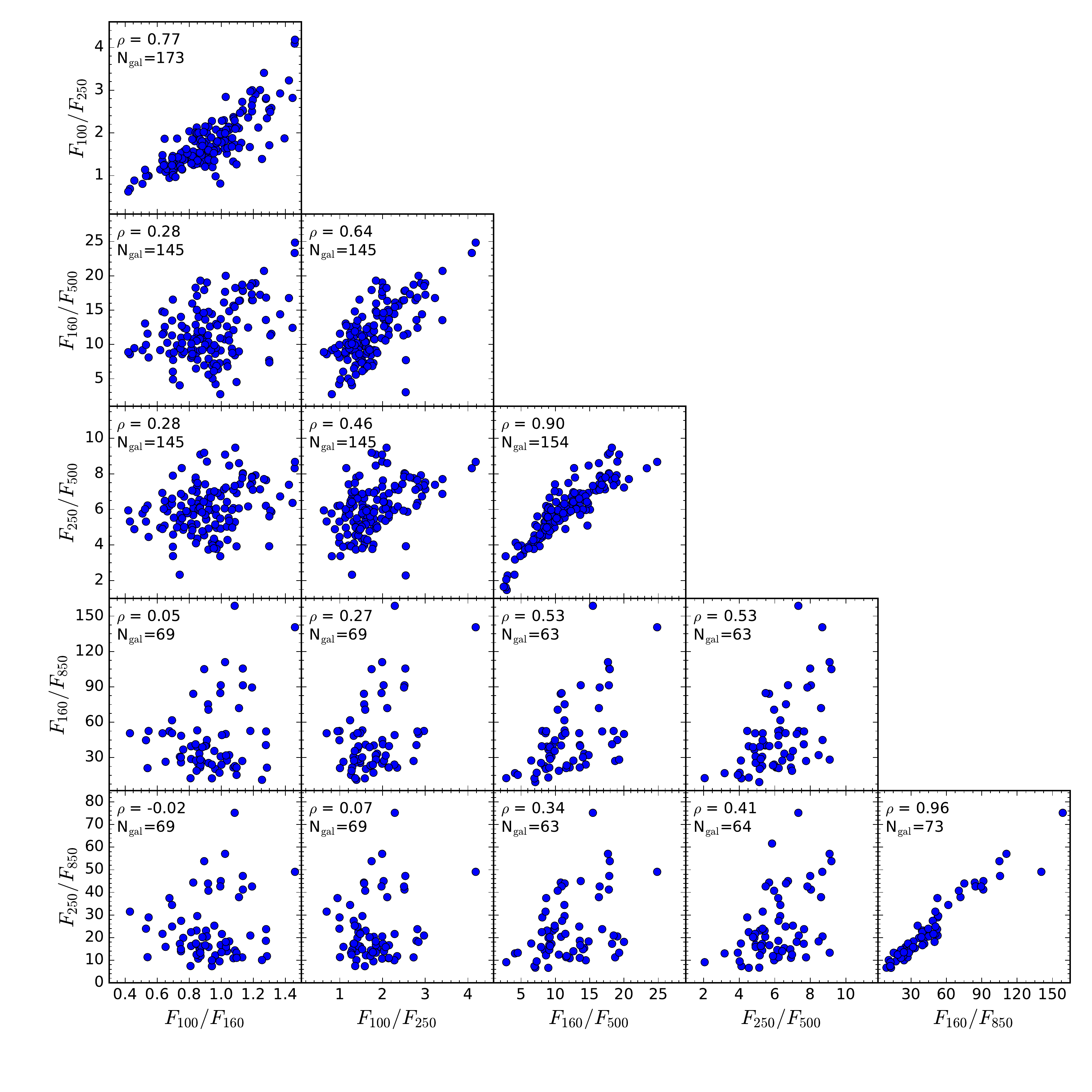}
\caption{Far-infrared/sub-mm colour-colour plots for selected colours, where each colour is the ratio of two 
flux-densitites in the wavelength range 100-850\micron. Galaxies are only included if each flux-density in the panel has a
signal-to-noise ratio greater than 3. The Spearman's rank correlation coefficient ($\rho$), and the number of galaxies in 
each panel is given in the top
left corner of each panel.}
\label{fig:FIRcolour-colour} 
\end{figure*}

Figure~\ref{fig:FIRcolour-colour} shows a grid of the FIR/sub-mm colour ratios (similar to those derived for the HRS in \citealt{Boselli2012} and \citealt{Cortese2014}),
the best correlations between FIR colours is seen for indicators that trace the dust emissivity index
(i.e., longer wavelengths). The highest correlation is seen between the $F_{250}/F_{850}$ and $F_{160}/F_{850}$,
with the Spearman rank coefficient ($\rho$) equal to 0.96.
This shows the advantage of having the longer 850\micron\ SCUBA-2 data to constrain the Rayleigh-Jeans tail, as using
only the SPIRE wavelengths $\rho = 0.80$,for the $F_{250}/F_{500}$ and $F_{250}/F_{350}$ colour ratio.
The two shorter colour-ratios ($F_{100}/F_{250}$ and $F_{100}/F_{160}$) also show a very good correlation
($\rho = 0.77$), probably
as both are indicative of dust temperature.

Similar correlations to those shown in Figure~\ref{fig:FIRcolour-colour} were found by \citet{Boselli2012}, for
the HRS \citep[][]{Boselli2010b}, with their best correlation ($\rho = 0.98$) between $F_{250}/F_{500}$
and $F_{250}/F_{350}$ . 
Overall, the \textit{Herschel} Reference Survey predominately had stronger
correlations than we find for JINGLE, although this could be explained due to the 
HRS targeting local galaxies with significantly higher global flux estimates, and deeper observations than \textit{H}-ATLAS.

\begin{figure*}
\centering
\includegraphics[width=0.75\textwidth]{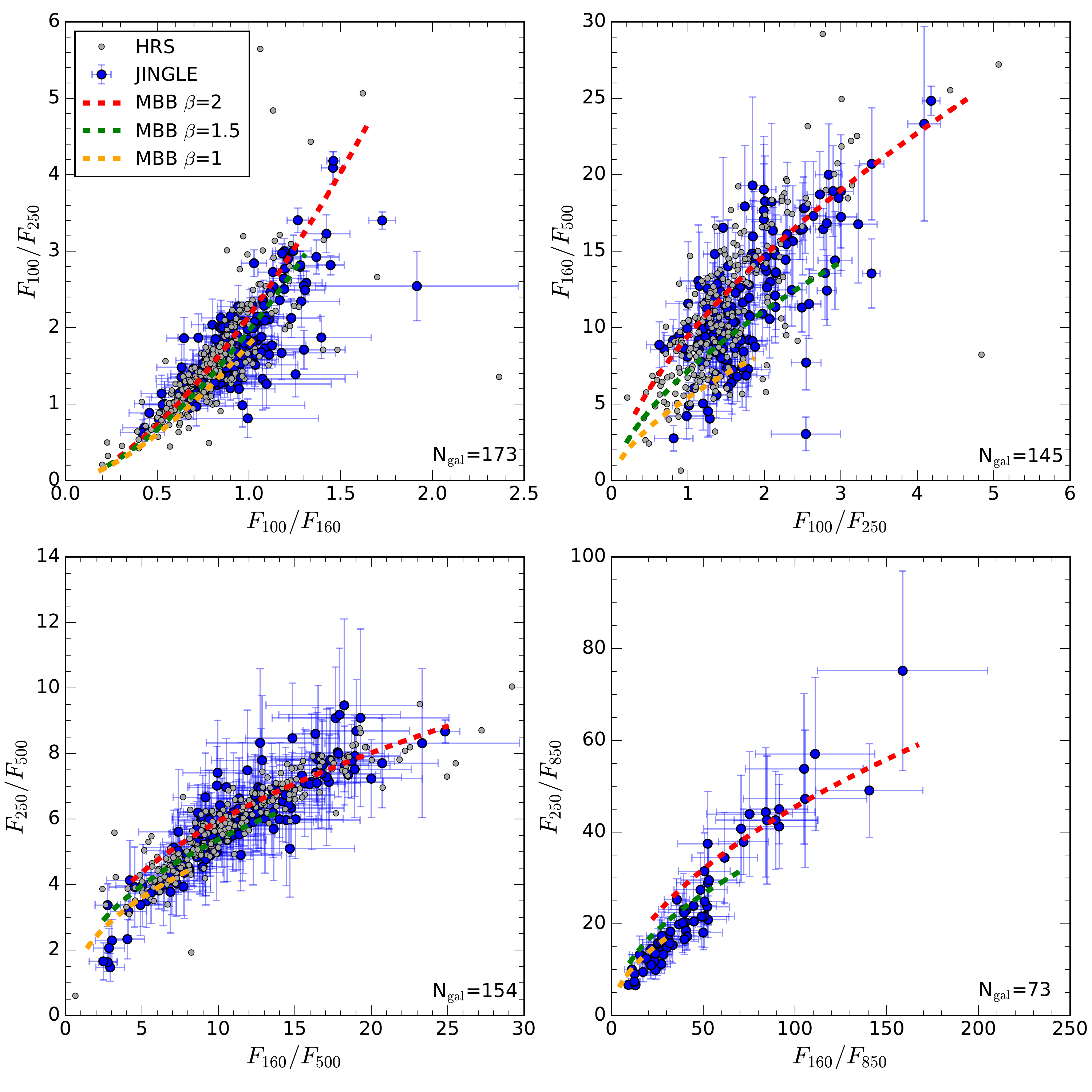}
\caption{The four far-infrared/sub-mm colour-colour plots from Figure~\ref{fig:FIRcolour-colour} with the strongest correlations. The
blue points with error bars are points with a signal-to-noise ratio greater than 3. The grey dots show the position
of detected galaxies from the HRS with no photometry flags in the catalogue. 
The red, green, and orange lines are the flux-density ratios 
of a single-modified blackbody between a temperature of 15 and 30\,K with a $\beta$ of 2, 1.5 and 1, respectively.}
\label{fig:FIRcolour-BBplot} 
\end{figure*}

In Figure~\ref{fig:FIRcolour-BBplot} we show the four colour-ratios from Figure~\ref{fig:FIRcolour-colour} with
the highest Spearman's rank correlation, including the uncertainties for each data point. For the colour-ratios
not including 850\micron\ we over-plot objects from the HRS in grey \citep{Boselli2012,Cortese2014}. Given the large difference in 
selection (the HRS is a local \textit{K}-band selected sample) compared to JINGLE, the distribution 
of the detected HRS galaxies in the colour ratio plots appears
very similar to that of JINGLE, suggesting the dust properties of both samples is broadly similar.
This is possibly due to the similar stellar masses of the samples, although the specific SFR is quite different (see JINGLE Paper IV, de Looze et al. \textit{in prep}).
The red, green and orange lines in Figure~\ref{fig:FIRcolour-BBplot} show the line produced for a 
single modified blackbody with a $\beta = 2$,
$\beta = 1.5$, and $\beta = 1$ over a temperature range of 15--30\,K respectively. The distribution of points in the
top-left panel is dominated by the temperature of the dust as shown by the similarity of the 
two blackbody lines. The blackbody lines show that a temperature range of 15--30\,K is 
sufficient to explain the vast majority of JINGLE and HRS objects (in agreement with JINGLE Paper V, Lamperti et al. \textit{in prep}). The top-right and
bottom left panels show the difficulty in making estimates of the dust emissivity index from \Hersc\ data alone
where both modified blackbody models with $\beta = 1.5$ and 2 lie within the scatter 
of the data for the sample. The bottom-right panel which shows the
$F_{250}/F_{850}$ versus $F_{160}/F_{850}$ shows the extra discrimination between models that the longer 
wavelength data provides, with the objects at high $F_{250}/F_{850}$ and $F_{160}/F_{850}$ appearing to lie on 
the $\beta = 2$ line, while surprisingly 
the majority of objects appear to fall below the $\beta = 1$ line.

To test the significance of our colour-plots to distinguish between the dust models, for each data point we calculated
the smallest `normalised' distance (i.e., the smallest distance between the model divided by the uncertainties). If the
model is an accurate description of the data, the distribution of the normalised distances should be a normal distribution.
We then use the Anderson-Darling test\footnote{We use the scipy implementation of the Anderson-Darling test \citep{scipy}.} to test
the null hypothesis that our distribution is drawn from the normal distribution. The $F_{160}/F_{500}$ versus $F_{100}/F_{250}$
reject the $\beta = 1.5$ and $\beta = 2.0$ models as the test statistic of 1.11 and 4.64 is above the critical level of 1.06 
and so the probability of our data being drawn from the distribution is $< 1\%$. The $F_{250}/F_{850}$ versus $F_{160}/F_{850}$
also rejects the the $\beta = 2.0$ model with a probability < 1\%, but is not as significant for $\beta = 1.5$ with $\sim$5\% 
probability the data is drawn from the distribution. However, the statistics with $F_{850}$ are likely to be an under-estimate
of the model rejection, as a plot of the residuals (i.e., data - model), shows a strong correlation ($\rho > 0.6$) and significant 
(p-values $<5 \times 10^{-9}$) as measured by the Spearman rank coefficient. 

Our results confirms the findings of \citet{Boselli2012} that a single modified blackbody model with
a constant $\beta$ cannot explain the distribution of points in both the \textit{Herschel} wavelengths
(i.e., the top-right panel in Figure~\ref{fig:FIRcolour-BBplot}), and the SCUBA-2 850\micron\ data.
While both single $\beta$ models are rejected by the \Hersc\ data, values between 1.5--2.0 would explain the majority
of data points. The bottom-left panel of Figure~\ref{fig:FIRcolour-BBplot}, shows many colour-ratios lie 
below the $\beta = 1.5$ line. It is possible to explain some of the systematic shift 
in the bottom-right panel by having an extreme calibration uncertainty for PACS and SPIRE, or by requiring 
a large colour-correction. However, a large calibration shift of the \Hersc\ data would lead to worse 
agreement in the other panels, and can only partly resolve the offset. The colour corrections for both PACS
and SPIRE are small ($\approx 3\%$) for a typical galaxy spectrum, and cannot resolve the offset. 
A potential solution is that 
a single dust component may not be a good model for a galaxy's spectrum and multiple temperatures along
the line-of-sight are required. \citet{Clark2015} performed an investigation of a blind dust survey of nearby
galaxies ($15 < D < 46$\,Mpc) using \textit{H}-ATLAS (the HAPLESS sample), and so has similar detection criteria 
to that of JINGLE. For their sample they found that they required a two temperature component distribution to
adequately describe their far-infrared fluxes, with many galaxies in their sample having a $\sim$9--15\,K cold
component. Using their best fitted parameters, a couple of the objects overlap with the extreme bottom left of 
the JINGLE distribution in a plot of 
$F_{250}/F_{850}$ verses $F_{160}/F_{850}$, the vast majority of HAPLESS galaxies lie between our single
modified blackbody with $\beta = 1.5$ and $\beta = 2.0$. However, the HAPLESS SED fits were performed assuming a 
$\beta = 2$ in the 60--500\micron\ wavelength range, and so extrapolating their models 
may not give the true HAPLESS flux.

A broken emissivity modified-blackbody model where the emissivity law changes $\beta$ value at a break wavelength 
\citep[where $\lambda_{\rm break} > $175\micron, see ][ for more details of the model]{Li2001,Gordon2014}, can be 
used to provide a better fit. In particular the cluster of objects with low 
$F_{250}/F_{850}$ versus $F_{160}/F_{850}$ values in Figure~\ref{fig:FIRcolour-BBplot} could be explained, although 
the break wavelength is at longer wavelengths (e.g., $\approx$500\micron). Even if such a model could explain
the excess at 850\micron, it is clear one dust model cannot explain the entire distribution of JINGLE galaxies.

Another possibility is that our signal-to-noise cut of 3 is artificially selecting galaxies that have
been boosted to high 850\micron\ flux densities. However, lowering the signal-to-noise cut to 2 results in the 
same observed discrepancy. 

In the next section we avoid this issue as for each galaxy we use the results
of fitting a modified blackbody model to each galaxy in JINGLE, as we have performed aperture matched 
photometry across all bands, even a low signal-to-noise 850\micron\ measurement can be useful to
constrain the fitted parameters.

\subsection{Far-infrared Colours as indicators of dust temperature and emissivity}
\label{sec:FIRdustBB}

\begin{figure}
\centering
\includegraphics[width=0.49\textwidth]{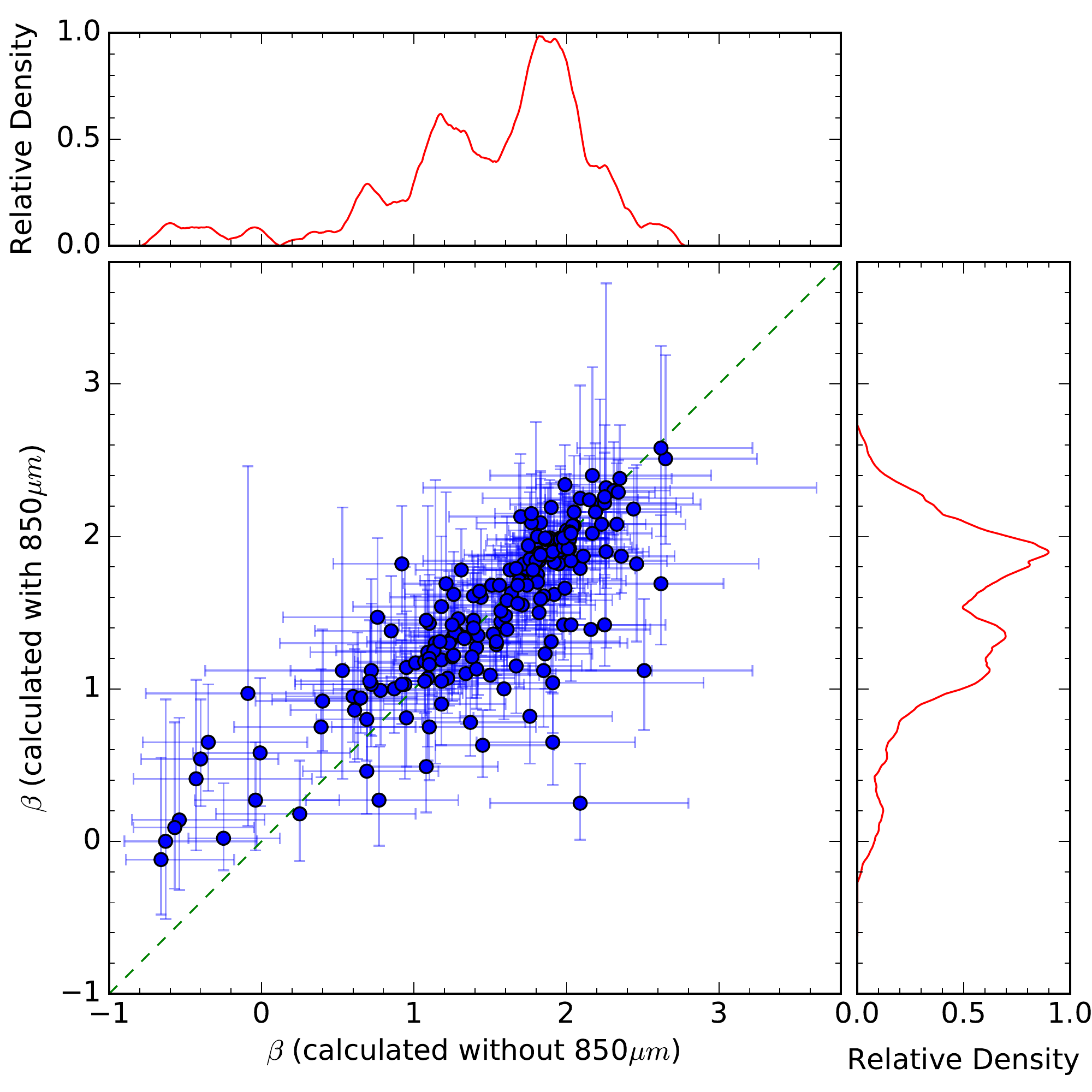}
\caption{Comparison of the emissivity index $\beta$ measured from the \Hersc\ bands and measured with the additional SCUBA-2 850\micron\ flux point. The green dashed line shows the one-to-one relation (i.e., when the two estimates
agree). The top and right panels show the KDE for each set of $\beta$ points, which clearly show how the 
850\micron\ flux changes the distribution of values measured.}
\label{fig:comp_beta} 
\end{figure}

We investigate the effect of including the 850\micron\ flux on the measurement of the effective emissivity 
index $\beta$. We perform the SED fitting with $\beta$ as a free parameter, and we compared $\beta$ measured 
from the fit with and without the 850\micron\ flux (Figure \ref{fig:comp_beta}). The points are scattered 
predominantly around the one-to-one relation. The largest discrepancies are 
for galaxies that have $\beta$ values $< 0$ when obtained using \Hersc\ data only, presumably due
to low signal-to-noise in the longer SPIRE wavelengths. When our 850\micron\ measurements are included
all but one of these objects have values above $\beta = 0$ (values of $\beta$ below 0 are not 
physically possible). 

The KDE panels in Figure~\ref{fig:comp_beta} show the distribution of $\beta$'s measured for the majority of
objects is similar before and after the 850\micron\ is included. However, 48\% of the sample have estimates of $\beta$ 
below 1.5, which may be an indication of a submm excess, 
(i.e., an excess emission at wavelengths $\geq 500$\micron\ with respect to what models would predict from
the 100--350\micron\ wavelength range).
Such an excess has been observed in dwarf 
galaxies \citep{Lisenfeld2002, Galliano2003}, in late-type 
galaxies \citep{Dumke2004, Bendo2006, Galametz2009, Relano2018}, as well as in the Magellanic Clouds 
\citep{Israel2010, Bot2010b, Planck2011, Gordon2014}. Several explanations have been proposed including a
very cold dust component ($< 10$\,K), spinning dust \citep{Anderson1993,Draine1998a,Draine1998b}, 
the Two-Level-System of amorphous dust \citep{Meny2007}, or a broken dust
emissivity law model \citep[where the emissivity changes at a `break wavelength'][]{Li2001}. Explanations invoking
very cold dust are problematic as they lead to very large estimated dust masses. In the LMC \citet{Gordon2014}
found the broken emissivity law model had the lowest residuals. These models will be investigated in 
detail in Paper V (Lamperti et al. \textit{in prep}).

\begin{figure*}
\centering
\includegraphics[width=0.8\textwidth, clip=True, trim=0mm 2mm 0mm 0mm]{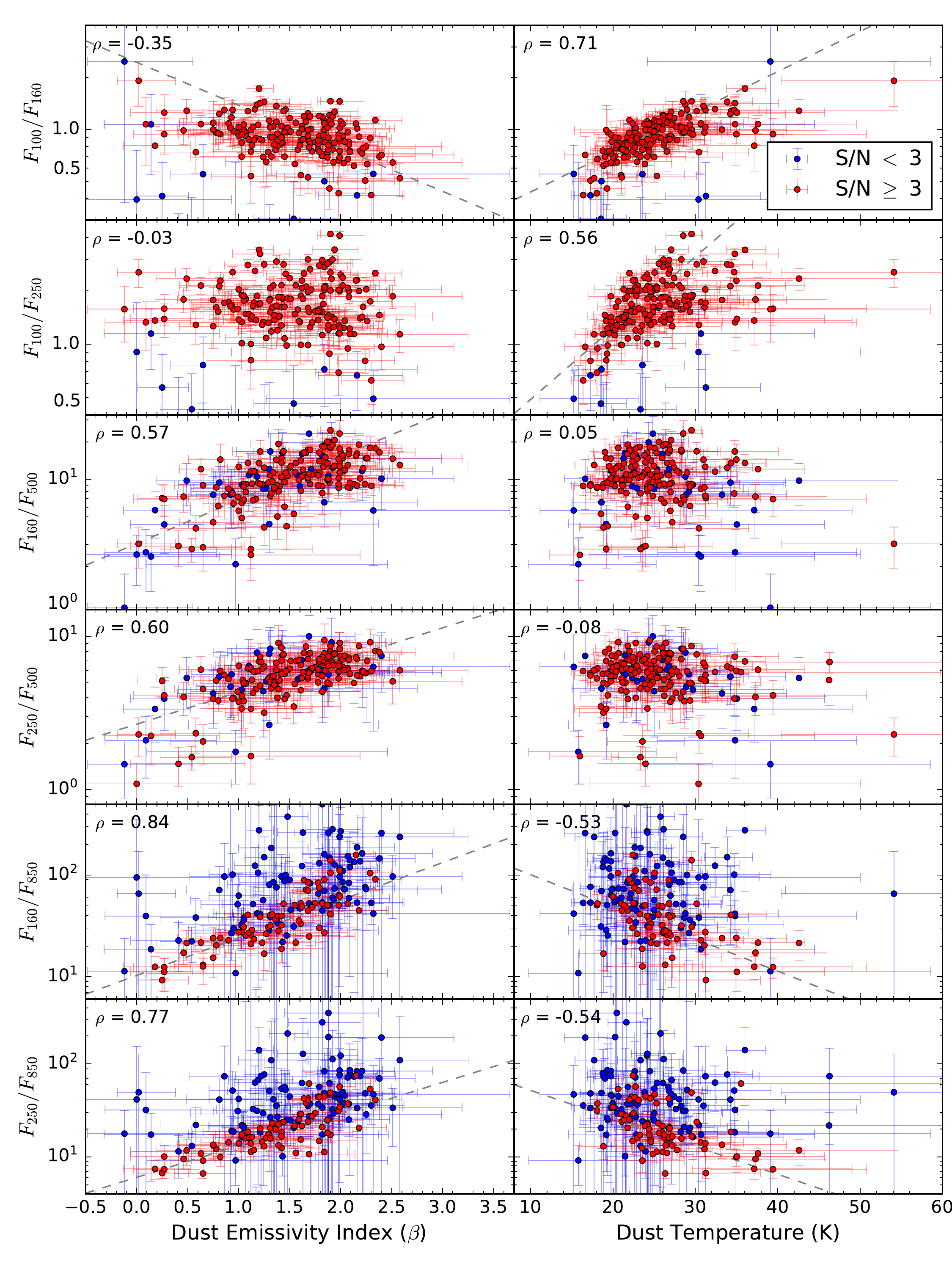}
\caption{Correlation between flux ratios and the dust emissivity $\beta$ (left) and dust temperature (right). In blue are shown 
points for which one of the flux measurements has high uncertainties (S/N < 3). The Pearson correlation coefficients 
$\rho$ for the objects with a signal-to-noise greater than 3 is shown in the top-left of each panel. 
The flux ratios $F_{100}/F_{160}$ and $F_{100}/F_{250}$ have the strongest correlation with the 
dust temperature ($\rho= 0.68$ and $\rho = 0.58$, respectively), although there are also negative correlations with the 
$F_{160}/F_{850}$ and $F_{250}/F_{850}$. 
The flux ratio $F_{160}/F_{850}$ and $F_{250}/F_{850}$ show the best correlation with $\beta$ 
(with $\rho= 0.82$ and 0.76, respectively). The grey lines show the best fit models specified in Table~\ref{tab:Fratio_corr}. 
Due to the large scatter of the lower signal-to-noise data points,
a few objects may lie outside the plotted range.
}
\label{fig:Fratio_vs_T_beta} 
\end{figure*}

The FIR colours that we have been studying can be used as an indicator of the cold dust temperature 
$T_{\text{dust}}$ and emissivity index $\beta$ 
\citep[e.g.][]{Boselli2010a, Boselli2012, Dale2012, Bendo2012, Galametz2010, Remy-Ruyer2013, Cortese2014}. 
We investigate which infra-red continuum flux ratios have the strongest correlation with 
$T_{\text{dust}}$ and $\beta$, measured from a modified blackbody fit including the 850\micron\ 
fluxes (see Figure \ref{fig:Fratio_vs_T_beta}).
This can be useful for surveys that do not have the wavelength coverage to do full SED fitting.

Figure~\ref{fig:Fratio_vs_T_beta}, shows the same FIR/sub-mm colour ratios versus our calculated dust emissivity
index and dust temperature. In each panel we compute the Pearson correlation coefficient ($\rho$), for all objects
with a signal-to-noise ratio greater than 3. The flux ratios $F_{100}/F_{160}$ and $F_{100}/F_{250}$ 
have the strongest correlations with the dust temperature ($\rho= 0.68$ and 0.58, respectively). These 
flux bands are sampling the peak of the SED, which for a typical dust temperature between 10--30\,K is 
in the wavelength range 90-250\micron. There are also fairly strong negative correlations with $\rho = -0.53$ and -0.54,
between dust temperature and the $F_{160}/F_{850}$ or $F_{250}/F_{850}$, respectively. We do not investigate
these further, but these may arise due to a temperature-$\beta$ degeneracy (for more discussion
see Paper V Lamperti et al. \textit{in prep}).

The flux ratio $F_{160}/F_{850}$ has the strongest correlation with the emissivity index $\beta$ ($\rho=0.82$), 
closely followed by the $F_{250}/F_{850}$ with $\rho = 0.76$. 
These colour-ratios are sampling the Rayleigh-Jeans tail of the sub-millimeter SED, and therefore are a good 
proxy of $\beta$. We fit the distribution of FIR/submm colour (with signal-noise ratios greater than 3) 
to either dust temperature or $\beta$, assuming
this model:
\begin{equation} 
\label{equ:ratioModel}
F_{i} / F_{j} = \gamma 10^{\alpha X}
\end{equation}
where $F_{i}$ and $F_{j}$ are the fluxes in the FIR/submm bands, $\alpha$ and $\gamma$ are the model parameters
and X is either dust temperature or $\beta$. We fit this model rather than a straight 
line to $\log_{10}{\left( F_i/F_j \right)}$ as
the flux uncertainties are assumed Gaussian in the linear regime. We perform the fits using the 
pyMC3 package \citep{Salvatier2016}, and incorporate uncertainties in both FIR/submm colour and dust model
parameters. Table~\ref{tab:Fratio_corr} shows the correlation coefficients, and the fitted model parameters 
to relate the FIR/submm colours with $\beta$ or dust temperature. These fits can be used to estimate 
$T_{\text{dust}}$ and $\beta$ if only two flux points are available.

For comparison, in Table~\ref{tab:Fratio_corr} we also include the correlation coefficients and fits possible 
with only \textit{Herschel} data. For estimates of $\beta$ we note the correlation is better if longer wavelength
information is available (e.g., $\rho = 0.54$ for $F_{250}/F_{500}$ versus $\rho = 0.79$ for $F_{250}/F_{850}$).

\begin{table*}
\centering
\caption{Correlation between flux ratios and dust properties (temperature and $\beta$).}
\label{tab:Fratio_corr}
\begin{tabular}{ccccccc} 
\hline\hline
Flux & \multicolumn{3}{c}{Dust Emissivity Index ($\beta$)} & \multicolumn{3}{c}{Dust Temperature} \\
Ratio & $\rho$ & $\alpha$ & $\gamma$ & $\rho$ & $\alpha$ ($\rm K^{-1}$)& $\gamma$ \\
\hline
$F_{100}$/$F_{160}$ & -0.35 & -0.248 $\pm$  0.016 & 2.45 $\pm$ 0.15 & \ 0.71 & 0.0232 $\pm$ 0.0013 & 0.256 $\pm$ 0.019 \\
$F_{100}$/$F_{250}$ & -0.03 & - & - & \ 0.56 & 0.0402 $\pm$ 0.0190 & 0.194 $\pm$ 0.020 \\
$F_{160}$/$F_{500}$ & \ 0.57 &\ 0.349 $\pm$ 0.019 & 3.06 $\pm$ 0.22 & \ 0.05 & - & -\\
$F_{250}$/$F_{500}$ & \ 0.60 &\ 0.210 $\pm$ 0.013 & 2.67 $\pm$ 0.14 & -0.08 & - & -\\
$F_{160}$/$F_{850}$ & \ 0.84 &\ 0.371 $\pm$ 0.031 & 10.3 $\pm$ 1.01 &  -0.53 & -0.0317 $\pm$ 0.0036 & 210.0 $\pm$ 52.6\\
$F_{250}$/$F_{850}$ & \ 0.77 &\ 0.305 $\pm$ 0.030 & 6.02 $\pm$ 0.56 &  -0.54 & -0.0301 $\pm$ 0.0035 & 104.0 $\pm$ 25.4\\
\hline

\end{tabular}\\
\begin{flushleft}
\textbf{Notes.} The table provides the Spearman rank coefficient ($\rho$), and model parameters for our distributions of
FIR/submm colours versus fitted dust parameters. For the model parameters described in Equation~\ref{equ:ratioModel} 
the median of the posterior distribution is given, and its uncertainty estimated from the 16th
and 84th percentile (the uncertainties are to a good approximation symmetrical). We only provide model fits for the 
parameters for distributions with $\lvert \rho \rvert > 0.3$. 
\end{flushleft}
\end{table*}

\subsection{Far-infrared Colours Versus Galaxy Physical Parameters}
\label{sec:colourPhysical}

In this final section we investigate if there are any relationships between the FIR/sub-mm colours 
and a few physical parameters of the galaxy from the quantities calculated in \paperI. We look for correlations with star-formation rate (SFR), stellar mass (M$_*$), 
specific star-formation rate ($\rm SFR/M_*$), the surface density of SFR ($\Sigma$(SFR)) and the surface density
of stellar mass ($\Sigma_*$). The two surface densities are calculated based on taking the integrated quantity 
(i.e., SFR or M$_*$) and dividing by the elliptical area of the galaxy using the Petrosian radius (in kpc) and the
axis ratio of each galaxy. 

\begin{figure*}
\centering
\includegraphics[width=0.9\textwidth]{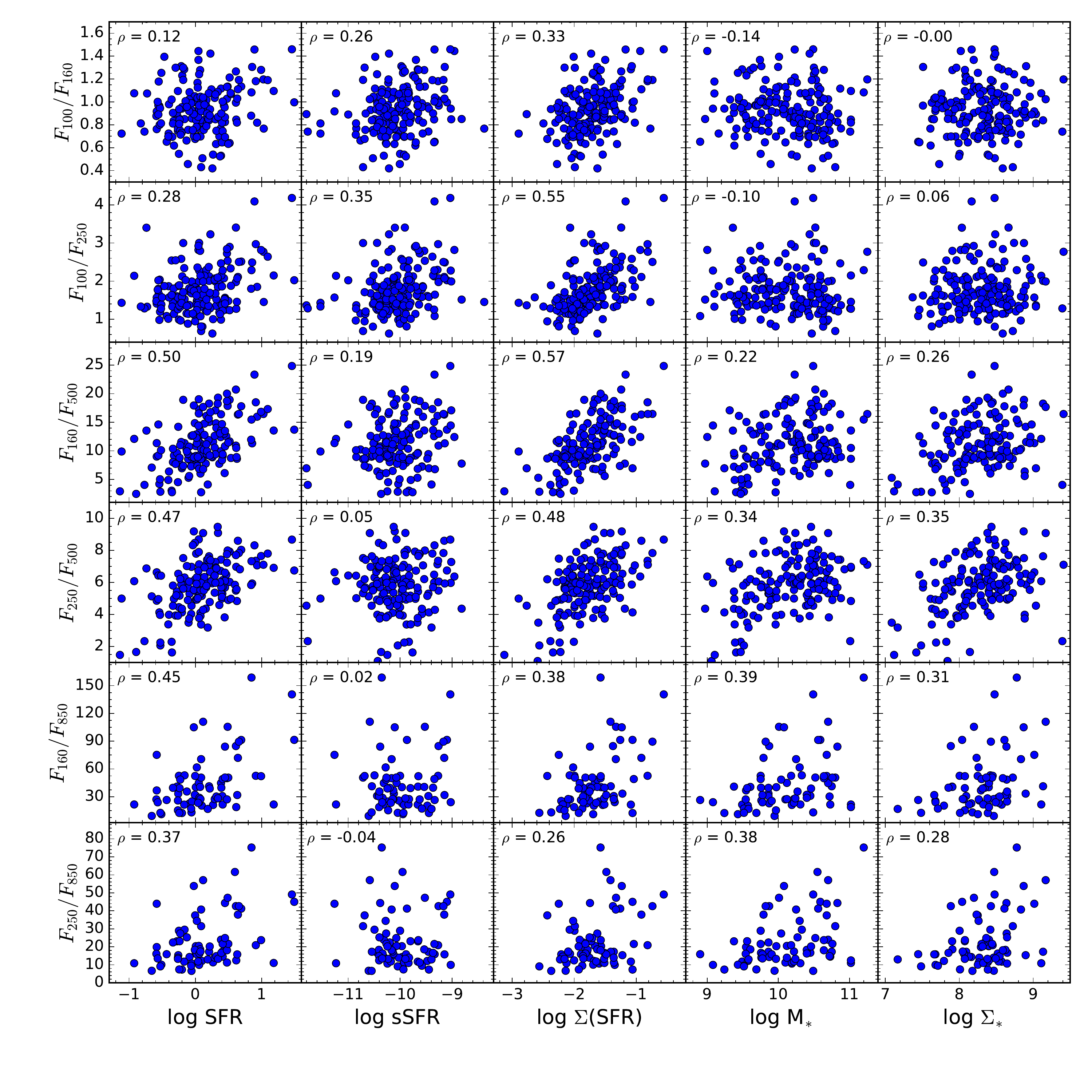}
\caption{FIR colours versus the SFR, sSFR, $\Sigma(SFR)$, M$_*$ and $\Sigma_*$, as defined in 
Section~\ref{sec:colourPhysical}. The Spearman rank correlation for each panel is shown in the top left
hand corner. Data points are included if the flux densities used to create each FIR/sub-mm colour have a
signal-to-noise of at least 3.}
\label{fig:FIRcolour-physicalParam} 
\end{figure*}

Figure~\ref{fig:FIRcolour-physicalParam} shows the FIR colours versus the various physical properties defined above.
The best correlation with a Spearman rank coefficient of $\rho = 0.57$ is between 
$\Sigma(SFR)$ and the $F_{160}/F_{500}$ (closely followed by the $F_{100}/F_{250}$ and $F_{250}/F_{500}$), and
similar, but reduced correlations are also seen with the SFR. 
Correlations with $\Sigma(SFR)$ and SFR would be expected due to the strength of the interstellar radiation field 
leading to an increase of the dust temperature, or equivalently far-infrared luminosity is used as a star formation
tracer \citep[e.g.,][]{Kennicutt1998}.
Surprisingly, from Section~\ref{sec:FIRdustBB} both the $F_{160}/F_{500}$ and $F_{250}/F_{500}$ colours are better
tracers of $\beta$, rather than dust temperature. This will be further investigated in Lamperti et al, in prep.
There are also some weak correlations with stellar mass and stellar mass surface density which again may be the
effect of increasing the interstellar radiation field and therefore increasing the dust temperature, however, our
results suggest for the majority of galaxies in JINGLE this is less significant than the heating from star formation.

\citet{Boselli2012} also compared their FIR colour-ratios for the HRS sample to a similar set of 
physical galaxy properties. For both samples a correlation is found between FIR colours at longer wavelength and 
the $\Sigma_*$ (or equivalently in the HRS study \textit{H}-band surface-brightness). Surprisingly, 
\citet{Boselli2012} did not find any correlations with SFR (although correlations with H$\alpha$ surface-density were found), 
this maybe due to the different method methods of measuring SFR (dust extinction-corrected H$\alpha$ or FUV emission versus panchromatic SED fitting for JINGLE), 
or the greater fraction of early-type galaxies in the HRS.

\section{Conclusions}

In this paper we have presented new SCUBA-2 data of the JINGLE sample which consists of 193 galaxies 
in the range 41-212\,Mpc. We described our data reduction tests and techniques which attempt to optimise
the SCUBA-2 data for the case of slightly extended galaxies, going beyond what is performed in the
standard SCUBA-2 pipeline. We investigate the optimum parameters, calibration and offset corrections
for our data. By incorporating data from \Hersc\ (both PACS and SPIRE) and WISE, we measure 
aperture-matched global fluxes across the dust SED. Our measurements attempt to account for
aperture corrections, effects from the pipeline filtering in the SCUBA-2 wavelengths, all sources
of noise in the FIR wavelengths (e.g., instrumental, confusion, cirrus, etc.), and the contamination
from the CO($J$=3-2) line.

Using the \Hersc\ data we find that the FIR/submm colours are similar to the 
HRS, and we find similar results to \citet{Boselli2012}, that the distribution on colour-colour
plots appears incompatible with a single modified blackbody with a constant $\beta$, but the majority 
of galaxies could be described with $\beta$ in the standard range of 1.5--2.0. By adding our 850\micron\
fluxes, we find that approximately half of the JINGLE objects
require a value of $\beta$ significantly lower than 1.5. Such low values are hard to reproduce with lower
temperatures and so possibly indicate a more complicated dust model like
those that have a broken-emissivity law is required. The distribution of JINGLE galaxies confirms one dust
model cannot explain the entire sample. These models applied to individual objects will be further 
investigated by Lamperti et al, in prep. who will use a Bayesian hierarchical fitting approach.

We found that the $F_{160}/F_{850}$ and the $F_{250}/F_{850}$ colours have the strongest correlation 
($\rho$ = 0.84 and 0.77, respectively) with
$\beta$ estimated from a single modified blackbody model. The dust temperature is better correlated with shorter
wavelength colours from \Hersc\ data, with the highest correlation found for the $F_{100}/F_{160}$ colour ($\rho$ = 0.71). 
We provide the fits to these plots to find an estimate of $\beta$ and temperature from a FIR/submm colour.

Finally, we investigate how the FIR/submm colours vary with different physical parameters presented in \paperI. 
The best correlation is between the FIR/submm colour and the 
surface-density of star-formation rate (closely followed by the total star-formation rate), indicating that for JINGLE galaxies dust heating is predominantly
due to young-stellar populations rather than older stellar populations. However, there is a significant but reduced
correlation with stellar surface-density, suggesting some dust heating from older stellar populations occurs in JINGLE galaxies.

\section*{Acknowledgements}

MWLS acknowledges support from the European Research Council (ERC) Forward Progress 7 (FP7)
project HELP. MWLS acknowledges funding from the UK Science and Technology Facilities Council 
consolidated grant ST/K000926/1.

E.B. acknowledges support from STFC, grant number ST/M001008/1.

I.D.L. gratefully acknowledges the support of the Research Foundation Flanders (FWO).

H.G. acknowledges support from the European Research Council (ERC) in the form of Consolidator grant
{\sc CosmicDust}.

The James Clerk Maxwell Telescope is operated by the East Asian Observatory on behalf of The 
National Astronomical Observatory of Japan; Academia Sinica Institute of Astronomy and Astrophysics; 
the Korea Astronomy and Space Science Institute; the Operation, Maintenance and Upgrading Fund for 
Astronomical Telescopes and Facility Instruments, budgeted from the Ministry of Finance (MOF) of 
China and administrated by the Chinese Academy of Sciences (CAS), as well as the National Key R\&D 
Program of China (No. 2017YFA0402700). Additional funding support is provided by the Science and 
Technology Facilities Council of the United Kingdom and participating universities in the United Kingdom and Canada.

We thank everyone involved with the {\it Herschel Space Observatory}.

SPIRE has been developed by a consortium of
institutes led by Cardiff University (UK) and including:
University of Lethbridge (Canada); NAOC (China); CEA, LAM (France); IFSI,
University of Padua (Italy); IAC (Spain); Stockholm Observatory (Sweden);
Imperial College London, RAL, UCL-MSSL, UKATC, University of Sussex (UK); and
Caltech, JPL, NHSC, University of Colorado (USA). This development has been
supported by national funding agencies: CSA (Canada); NAOC (China);
CEA, CNES, CNRS (France); ASI (Italy); MCINN (Spain); 
SNSB (Sweden); STFC, UKSA (UK); and NASA (USA). 

This research made use of Astropy, a community-developed core Python package for Astronomy (Astropy Collaboration, 2013).




\bibliographystyle{mnras}
\bibliography{jingleDR} 





%
%


\bsp	
\label{lastpage}
\end{document}